\documentclass[iop]{emulateapj}

\usepackage{graphicx}
\usepackage{amsmath}
\def\dlambda{$\lambda\lambda$}
\def\kms{km~s$^{-1}$} 

\slugcomment{Accepted for publication in the Astrophysical Journal on February 29, 2012}

\shorttitle{Late-Time Optical Emission From CCSNe}
\shortauthors{Milisavljevic et al.}

\begin{document}

\title{Late-Time Optical Emission From Core-Collapse Supernovae
}

\author{Dan Milisavljevic\altaffilmark{1}$^{\rm,}$\altaffilmark{5},
        Robert A.~Fesen\altaffilmark{1},
        Roger A.~Chevalier\altaffilmark{2},
        Robert P.~Kirshner\altaffilmark{3},
        Peter Challis\altaffilmark{3}, and 
        Massimo Turatto\altaffilmark{4} }

\altaffiltext{1}{6127 Wilder Lab, Department of Physics \& Astronomy, Dartmouth
                 College, Hanover, NH, 03755}
\altaffiltext{2}{P.O. Box 400325, University of Virginia, Charlottesville, VA
                 22904-4325}
\altaffiltext{3}{Harvard-Smithsonian Center for Astrophysics, 60 Garden Street,
                 Cambridge, MA, 02138} 
\altaffiltext{4}{Osservatorio Astronomico di Padova, vicolo dell'Osservatorio 5, 
                 I-35122, Padova, Italy}
\altaffiltext{5}{Now at: Harvard-Smithsonian Center for Astrophysics,
                  60 Garden Street, Cambridge, MA, 02138; dmilisav@cfa.harvard.edu}

\begin{abstract}

  Ground-based optical spectra and {\it Hubble Space Telescope} images of ten
core-collapse supernovae (CCSNe) obtained several years to decades after
outburst are analyzed with the aim of understanding the general properties of their
late-time emissions. New observations of SN~1957D, 1970G, 1980K, and 1993J are
included as part of the study. Blueshifted line emissions in oxygen and/or
hydrogen with conspicuous line substructure are a common and long-lasting
phenomenon in the late-time spectra.  Followed through multiple epochs, changes
in the relative strengths and velocity widths of the emission lines are
consistent with expectations for emissions produced by interaction between SN
ejecta and the progenitor star's circumstellar material.  The most distinct
trend is an increase in the strength of
[\ion{O}{3}]/([\ion{O}{1}]+[\ion{O}{2}]) with age, and a decline in
H$\alpha$/([\ion{O}{1}]+[\ion{O}{2}]) which is broadly consistent with the view
that the reverse shock has passed through the H envelope of the ejecta in many
of these objects. We also present a spatially integrated spectrum of the young
Galactic supernova remnant Cassiopeia~A (Cas~A).  Similarities observed between
the emission line profiles of the $\approx$330 yr old Cas~A remnant and
decades old CCSNe suggest that observed emission line asymmetry in evolved CCSN spectra
may be associated with dust in the ejecta, and that minor peak substructure
typically interpreted as `clumps' or `blobs' of ejecta may instead be linked
with large-scale rings of SN debris.

% The kinematic properties of Cas~A's ejecta are mapped with
%   the spectrum's blueshifted line profiles to connect observed
%   emission asymmetry with internal absorption due to dust and
%   multi-peaked line substructure with its multi-ringed O- and S-rich
%   ejecta. 

\end{abstract}


\section{Introduction}
\label{sec:Intro}

Optical spectra of core-collapse supernovae (CCSNe) beyond a couple years after
maximum light are difficult to obtain due to their increasing faintness with
time and thus are relatively rare.  Given that the majority of CCSNe occur at
distances $>10$ Mpc and fade at least eight magnitudes below peak brightness
within their first two years \citep{Kirshner90}, observations have been largely
limited to the first 700 days or so after maximum light when they are at
apparent magnitudes $\la$ 20.

However, in a handful of cases it has been possible to monitor CCSNe several
years or even decades post-outburst.  This may be because of a fortuitously
nearby distance, such as SN 1987A in the Large Magellanic Cloud (D $\sim$ 50
kpc; \citealt{Kunkel87,Feast99,vanLeeuwen07}), or exceptional circumstances
wherein some late-time energy source maintains optical luminosity at observable
levels. This latter scenario was first recognized in the late 1980s with the
optical re-detections of SN~1980K \citep{Fesen88} and SN~1957D
\citep{Long89,Turatto89}.

Of the various late-time mechanisms theorized to power CCSN emission at epochs
$>$ 2 yr, the most common and best understood process is the forward shock front
and SN ejecta interaction with surrounding circumstellar material (CSM) shed
from the progenitor star (see \citealt{Chevalier03,Chevalier06} and references
therein).  SN--CSM interaction emits across a wide spectral band spanning radio
to X-ray.  Optical emissions largely originate from a reverse shock that
propagates upstream into outward expanding ejecta that gets heated and ionized
\citep{ChevFran94}.  The most dominant emission lines are [\ion{O}{1}] \dlambda
6300, 6364, [\ion{O}{2}] \dlambda 7319, 7330, [\ion{O}{3}] \dlambda 4959, 5007,
and H$\alpha$ with broad linewidths $\ga 2000$ \kms.  Other proposed late-time
mechanisms include pulsar/magnetar interaction with expanding SN gas
\citep{ChevFran92,Kasen10,Woosley10} or accretion onto a black-hole remnant
\citep{Patnaude11}.

\begin{deluxetable*}{llrrlccc}
\footnotesize
\centering
\tablecaption{Summary of {\it HST} Images}
\tablecolumns{8}
\tablewidth{0pt}
\tablehead{\colhead{Inst.}                        &
           \colhead{Filter}                       &
           \colhead{$\lambda _{\rm cen}$}         &
           \colhead{$\lambda _{\rm FWHM}$}        &
           \colhead{Descriptive}                  &
           \colhead{Date}                         &
           \colhead{Exp Time}                     &
           \colhead{Program No./PI}               \\
           \colhead{}                             &
           \colhead{}                             &
           \colhead{(\AA)}                        &
           \colhead{(\AA)}                        &
           \colhead{Notes}                             &
           \colhead{(UT)}                         &
           \colhead{(s)}                          &
           \colhead{}                             }
\startdata
\multicolumn{8}{c}{SN 1957D}\\
\hline\noalign{\smallskip}

 WFC3  & F336W & 3375 & 550  & Johnson $U$             	           & 17 Mar 2010 & 2560 & 11360/R.\ O'Connell \\
       & F438W & 4320 & 695  & Johnson $B$             	           & 17 Mar 2010 & 1800 & \\
       & F502N & 5013 & 47   & [\ion{O}{3}] 	           & 19 Mar 2010 & 2484 & \\
       & F547M & 5475 & 710  & Continuum                 & 20 Mar 2010 & 1203 & \\
       & F657N & 6573 & 94   & H$\alpha$ + [\ion{N}{2}]            & 19 Mar 2010 & 1484 & \\
       & F673N & 6731 & 77   & [\ion{S}{2}]             & 17 Mar 2010 & 1770 & \\

\hline\noalign{\smallskip}
\multicolumn{8}{c}{SN 1970G}\\
\hline\noalign{\smallskip}

 WFPC2  & F606W & 5997  & 1502 & Wide-$V$  & 21 Apr 1998 & 600  & 6713/W.\ Sparks \\
        & F656N & 6564  & 22   & H$\alpha$ & 21 Apr 1998 & 1600 & \\

\hline\noalign{\smallskip}
\multicolumn{8}{c}{SN 1980K}\\
\hline\noalign{\smallskip}

WFPC2   & F606W & 5997 & 1502 & Wide-$V$ &    19 Jan 2008 & 1600 & 11229/M.\ Meixner \\
        & F814W & 7940 & 1531 & Johnson $I$ & 19 Jan 2008 & 1600 & 
\enddata
\label{tab:hst}
\end{deluxetable*}

Studies of late-time optical emissions can yield kinematic and
chemical information about the SN ejecta and probe the mass-loss
history and evolutionary status of the progenitor stars
\citep{Leibundgut91,Fesen99}.  The evolution of line widths and
differences between the relative line strengths can distinguish
between late-time mechanisms and be important diagnostics of the
ejecta structure.

Here we present optical images and spectra of ten CCSNe observed
during the relatively unexplored late-time transition phase between SN
outburst and SN remnant formation.  In Sections \ref{sec:Obs} and
\ref{sec:Results} we present and briefly discuss high-resolution
images and new low-resolution optical spectra of SN~1957D, SN~1970G,
SN~1980K, and SN~1993J along with archival late-time spectra of six
other CCSNe. Common properties observed in these data are reviewed in
Section \ref{sec:Discussion} and then interpreted in the context of a
SN--CSM interaction model in Section \ref{sec:Models}. Further
investigation of these spectra follow from comparing them to a
spatially integrated spectrum of the young Galactic supernova remnant
(SNR) Cassiopeia~A in Section \ref{sec:CasA}, and a summary of our
findings is given in Section~\ref{sec:Conclusions}.

\section{Observations}
\label{sec:Obs}

\subsection{Images}

High-resolution images obtained with the {\sl Hubble Space Telescope}
({\sl HST}) covering the sites of SN~1957D, 1970G, and 1980K were
retrieved from the Multimission Archive at Space Telescope (MAST) and
Hubble Legacy Archive (HLA) maintained by the Space Telescope Science
Institute (STScI) to examine the supernovae and their local
environments. Table~\ref{tab:hst} lists the instruments, filter
passbands with notes about their emission line sensitivies, dates,
exposure times and program details of the observations. All images
retrieved from MAST were manually co-added and cleaned of cosmic rays
using the {\tt crrej} and {\tt multidrizzle} tasks in
IRAF/PyRAF\footnote{IRAF is distributed by the National Optical
  Astronomy Observatories, which are operated by the Association of
  Universities for Research in Astronomy, Inc., under cooperative
  agreement with the National Science Foundation. PyRAF is a product
  of the Space Telescope Science Institute, which is operated by AURA
  for NASA.}, whereas those retrieved from the HLA were automatically
drizzled by the archive through pipelined software.

\subsection{Spectra}

Late-time, low-dispersion optical spectra of four CCSNe were obtained using a
variety of telescopes and instrumental setups. Below we describe the details of
these observations.

Spectra of SN 1957D were obtained on 2001 July 27 with the FOcal Reducer/low
dispersion Spectrograph 1 (FORS1) at ESO/VLT at Paranal, Chile. The CCD detector
has a scale of 0$\farcs$2 pixel$^{-1}$, and the GRIS300V grism with a GG435
filter was used with a 1$''$ slit to produce spectra ranging from $4400-8700$
\AA\ with full-width-at-half-maximum (FWHM) resolution of approximately 10
\AA. Total exposure time was 1800~s obtained at an airmass of 1.7. An
atmospheric dispersion corrector minimized potential effects of atmospheric
refraction.

Observations of SN 1970G were obtained on 2010 March 12 with the 2.4~m Hiltner
telescope at MDM Observatory on Kitt Peak, Arizona.  A Boller \& Chivens CCD
spectrograph (CCDS) was used with a north-south 1.5$\arcsec \times$ 5$\arcmin$
slit, a 150 lines mm$^{-1}$ 4700 \AA\ blaze grating, and a LG400 filter to block
contaminating second-order light.  Exposures totaling $2 \times 3600$~s obtained
at culmination at airmasses $<$ 1.1 were combined. Resulting spectra spanned
$4300-7900$ \AA\ with resolution of 11 \AA. Conditions were mostly
photometric with the seeing around 1$\farcs$2. 

The 2.4 m Hiltner telescope at MDM was also used to obtain spectra of SN~1980K
on 2010 October 9.  The Mark III Spectrograph with a SITe 1024 $\times$ 1024 CCD
detector was used.  A 1$\farcs$2 $\times$ 4$\farcm$5 slit and a 300 lines
mm$^{-1}$ 5400~\AA\ blaze grism were employed. A total of 2 $\times$ 3000 s
exposures were obtained at culmination around an airmass of 1.1 with a spectral
resolution of 7 \AA. Conditions were photometric with sub-arcsecond seeing.

Spectra of SN 1993J were obtained on 2009 December 9 with the 6.5~m
MMT at Mt.\ Hopkins in Arizona using the HECTOSPEC optical fiber fed
spectrograph.  These observations were part of a survey of supernova
remnants of M81. Spectra from the 1$\farcs$5 diameter
fibers cover the wavelength range of $3700-9200$~\AA\ with FWHM
resolution of 5~\AA. Observations were obtained at an airmass of 1.3
with an atmospheric dispersion corrector in place and the total exposure time
was 3600~s.

These spectra were reduced and calibrated employing standard techniques in
IRAF and our own IDL routines (see \citealt{Matheson08}).
Cosmic rays and obvious cosmetic defects have been removed from all spectra.
Wavelengths were checked against night sky emission lines. Flux calibrations
were from observations of \citet{Stone77} and \citet{Massey90} standard stars.

\begin{figure*}[!htp]
\centering
\includegraphics[width=0.85\linewidth]{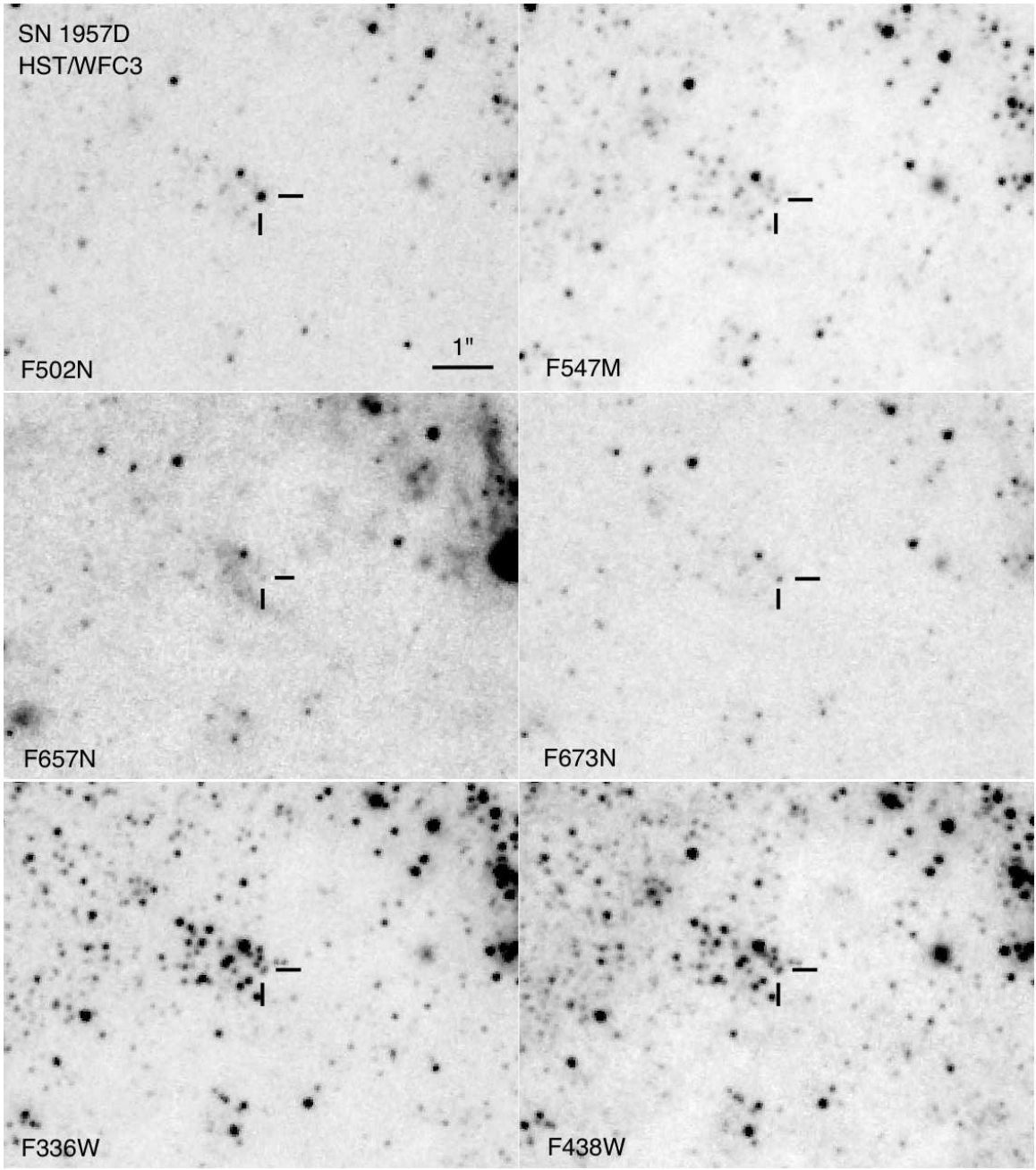}

\caption{{\sl HST} WFC3/UVIS images of the region around SN 1957D obtained
2010 March. The location of the SN is marked. For these images and all those
that follow, North is up and East is to the left.}

\label{fig:sn57Dimages}
\end{figure*}

\section{Late-Time CCSN Spectra \& Images}
\label{sec:Results}

\subsection{SN 1957D}

SN~1957D in M83 (D $\sim$ 4.6 Mpc; \citealt{Saha06}) was discovered by
H.\ Gates on 1957 December 28 when the SN was well past maximum light
in a spiral arm about $3\arcmin$ NNE of the galaxy nucleus
\citep{Gates58}.  The SN was followed photometrically on 1958 January
16, 30 and February 14 when its photographic magnitude was $16.7$. No
spectrum was obtained during this early period.

Approximately 25 yr after outburst, the SN was recovered as a radio
source \citep{Cowan82,Pennington83}. Subsequent observations by
\citet{Long89} and \citet{Turatto89} showed that its optical spectra
were dominated by broad [\ion{O}{3}] \dlambda 4959, 5007
lines. Additional follow-up observations by \citet{Long92} showed that
the optical flux had decreased by at least a factor of five between
1987 and 1991, and \citet{Cappellaro95} confirmed the rapid
decline.  \citet{Pennington82} suggested that the SN was produced by a
massive Population I progenitor based on integrated colors in the
surrounding region, corroborating suspicion that this was a Type II or
Ib/c event \citep{Long89}.

In Figure~\ref{fig:sn57Dimages}, we show high-resolution {\sl HST}
images of the region around SN 1957D which provide the first clear
look at the supernova's neighboring stellar environment.  The O-rich
SN is bright and unresolved in the F502N image. We estimate an
[\ion{O}{3}] $\lambda$5007 flux of $3.4 \pm 0.2 \times 10^{-16}$ erg
s$^{-1}$ cm$^{-2}$, which is a decline from April 1991 when it was
$1.8 \times 10^{-15}$ erg s$^{-1}$ cm$^{-2}$ \citep{Long92}. The SN is
marginally detected in the H$\alpha$-sensitive F657N image where we
see its location just outside a nearby \ion{H}{2} region.  The SN is
also visible in the [\ion{S}{2}]-sensitive F673N image with emission of the
order of $2 \times 10^{-17}$ erg s$^{-1}$ cm$^{-2}$. Emission is
stronger in the F673N image than in the F657N image, suggesting
[\ion{S}{2}] \dlambda 6716, 6731 line emission associated with shocked
SN ejecta.

The F336W and F438W images indicate that the remnant is projected
along the outskirts of a cluster of blue stars with dimensions of
approximately 1$\arcsec$ $\times$ 1$\farcs$5, or 25 $\times$ 35 (d/4.6
Mpc) pc.  Emission in these filter images, bright in the $U$ and $B$
passbands, is seen at the location of the SN, likely associated with
nearby massive stars. The stellar environment of SN~1957D resembles
the luminous SN remnant in NGC 4449 as observed with {\sl HST} (SNR
4449-1; \citealt{Milisavljevic08a}). This is another O-rich, young
remnant with an age of $\sim 70$ yr estimated from the ratio of
anugular size
to expansion velocity \citep{Bietenholz10b}.

\begin{figure}[!htp]
\centering
\includegraphics[width=\linewidth]{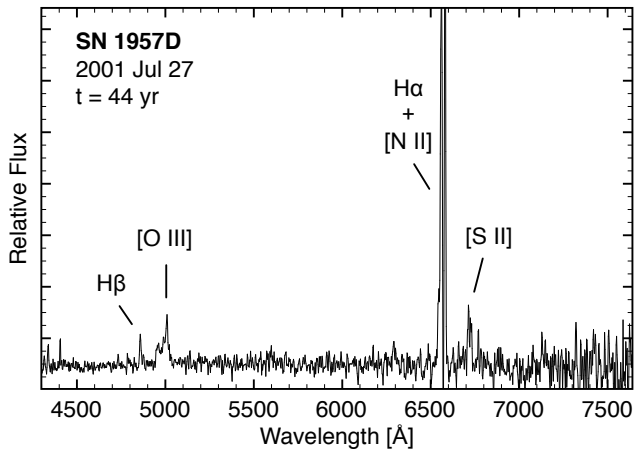}

\caption{Optical spectrum of SN~1957D obtained with the VLT on 2001
  July 27.  Broad [O~III] \dlambda 4959, 5007 emission originates from
  the SN, while the narrow H$\alpha$, [\ion{N}{2}] \dlambda 6548,
  6583, H$\beta$, and [S~II] \dlambda 6716, 6731 lines are associated
  with a nearby \ion{H}{2} region (see
  Figure~\ref{fig:sn57Dimages}). }

\label{fig:sn57D_fullspec}
\end{figure}

In Figure~\ref{fig:sn57D_fullspec}, we present a 2001 optical spectrum
of SN~1957D, some 44 yr after outburst. Wavelengths have been
corrected for the host galaxy's redshift of 513 \kms\ and a blue
continuum associated with background starlight has been subtracted
using a third order Chebyshev function.  Narrow H$\alpha$, H$\beta$,
[\ion{N}{2}] \dlambda 6548, 6583, and [\ion{S}{2}] \dlambda 6716, 6731
line emissions are associated with the nearby \ion{H}{2} region, and
the sharp cut in the H$\alpha$ profile is due to poor subtraction of
this emission. The dominant emission lines observed from the SN ejecta
are broad [\ion{O}{3}] \dlambda 4959, 5007.

\begin{figure}[!htp]
\centering
\includegraphics[width=0.95\linewidth]{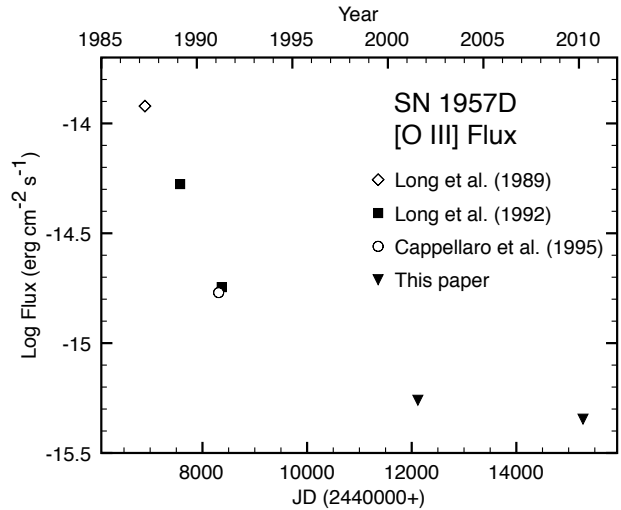}

\caption{History of reported [O~III] \dlambda 4959, 5007 fluxes for SN~1957D.
  All measurements are from the spectral lines except for the 2010.3 {\sl HST}
  observation which is from a narrow band [O~III] $\lambda$5007 image and has
  been corrected to compensate for the unsampled $\lambda$4959 line.}

\label{fig:sn57D_OIIIflux}
\end{figure}

\begin{figure}[!htp]
\centering
\includegraphics[width=0.75\linewidth]{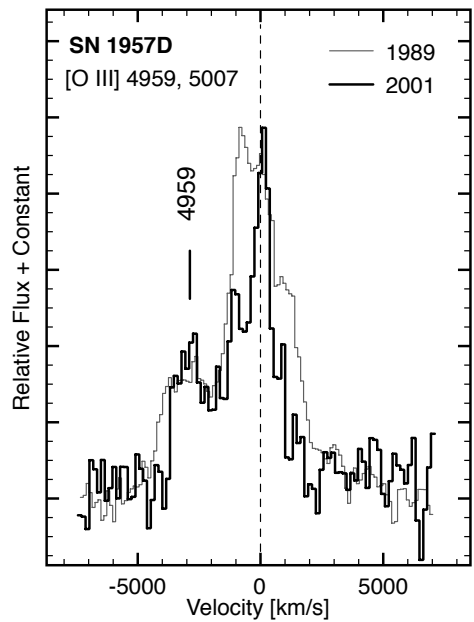}

\caption{Emission line profiles of SN~1957D's [O~III] \dlambda 4959, 5007
emission in 1989 and 2001. Velocities are with respect to 5007 \AA. }

\label{fig:sn57D_2epoch}
\end{figure}

The half-width-at-zero intensity (HWZI) of the [\ion{O}{3}] emission
is approximately 1000 \kms, confirmed by measuring both from 5007 \AA\
to the red and 4959 \AA\ toward the blue.  The measured [\ion{O}{3}]
\dlambda 4959, 5007 flux is $5.5 \pm 0.5 \times 10^{-16}$ erg s$^{-1}$
cm$^{-2}$, which is consistent with the decline between the
\citet{Long92} and the 2010 {\sl HST} WFC3 F502N image
measurements. In Figure~\ref{fig:sn57D_OIIIflux}, we plot the handful
of published [\ion{O}{3}] fluxes. A clear decrease in flux is seen
between previously reported values from \citet{Long92} and
\citet{Cappellaro95} and the 2001 and 2010 measurements presented
here. The last two measurements give the impression that the decline
in flux has slowed or even leveled out.

In Figure~\ref{fig:sn57D_2epoch}, we show a comparison of the 2001
July [\ion{O}{3}] line profile of SN~1957D to a
spectrum obtained in 1989 April \citep{Turatto89}. Some narrowing of
the emission line of order $-1000$ \kms\ may have occurred between the
ten years that separates the observations ($\sim$ 100 \kms\ yr$^{-1}$)
although the signal-to-noise (S/N) ratio prevents a firm conclusion. Our
1989 measurement is in agreement with \citet{Long92} who reported
[\ion{O}{3}] emission line velocities in excess of $\pm$2000 \kms\ in
their optical spectra taken at a similar epoch.

\begin{figure*}[!htp]
\centering
\includegraphics[width=0.97\linewidth]{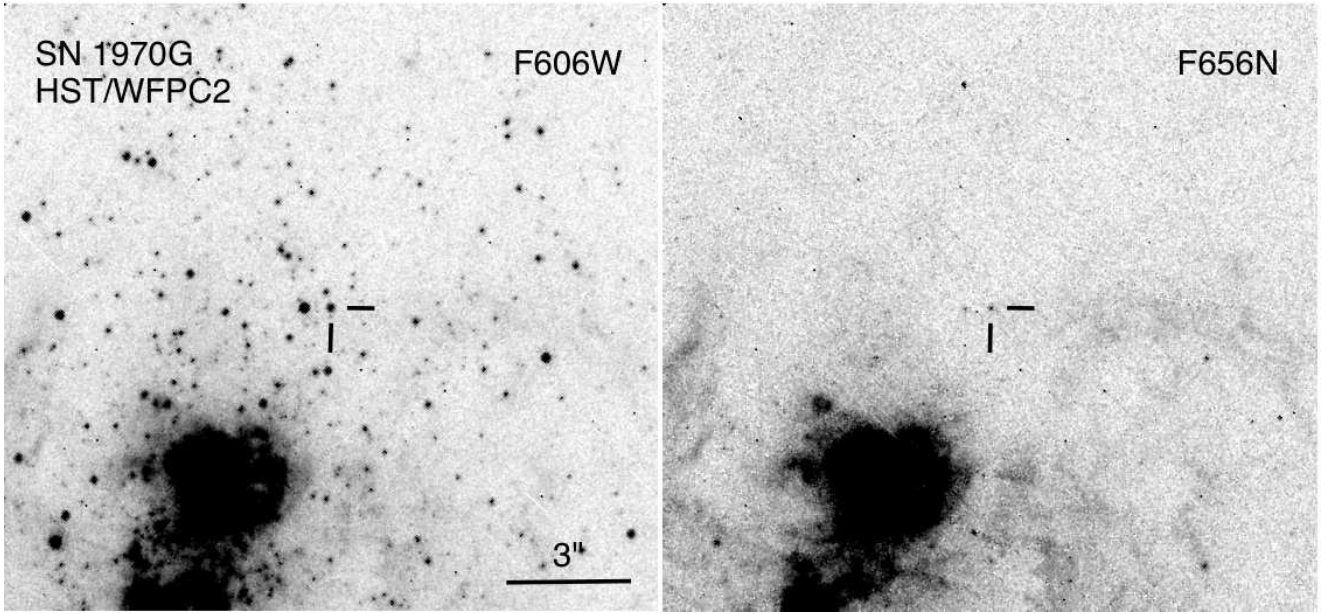}

\caption{{\sl HST} WFPC2 images of the region around SN 1970G obtained 1998
  April 21.  {\it Left}: F606W (Wide $V$) image in high contrast marking
  SN~1970G as one of the brightest objects in the field. {\it Right}: F656N
  image sensitive to H$\alpha$.}

\label{fig:sn70G_image}
\end{figure*}

\subsection{SN 1970G}

SN~1970G in M101 (D $\sim$ 6.7 Mpc; \citealt{Freedman01}) was discovered by
M.\ Lovas on 1970 July 30 \citep{Detre70} near maximum light ($B$ = 11.5;
\citealt{Winzer74,Barbon79}) along the NW boundary of the galaxy's large
\ion{H}{2} complex NGC~5455. Optical spectra showed Type II SN features with
broad H$\alpha$ emission \citep{Kirshner73,Kirshner74}.  It has been generally
classified as a Type IIL based on its linearly declining light curve
\citep{Barbon79,Young89}, but a brief plateau phase between day 30 and 50 and
optical spectra exhibiting prominent P-Cygni line absorption features may make
it a transitional object between IIP and IIL
\citep{Barbon79,Turatto90,Cappellaro91}.

\begin{figure}[!htp]
\centering
\includegraphics[width=0.95\linewidth]{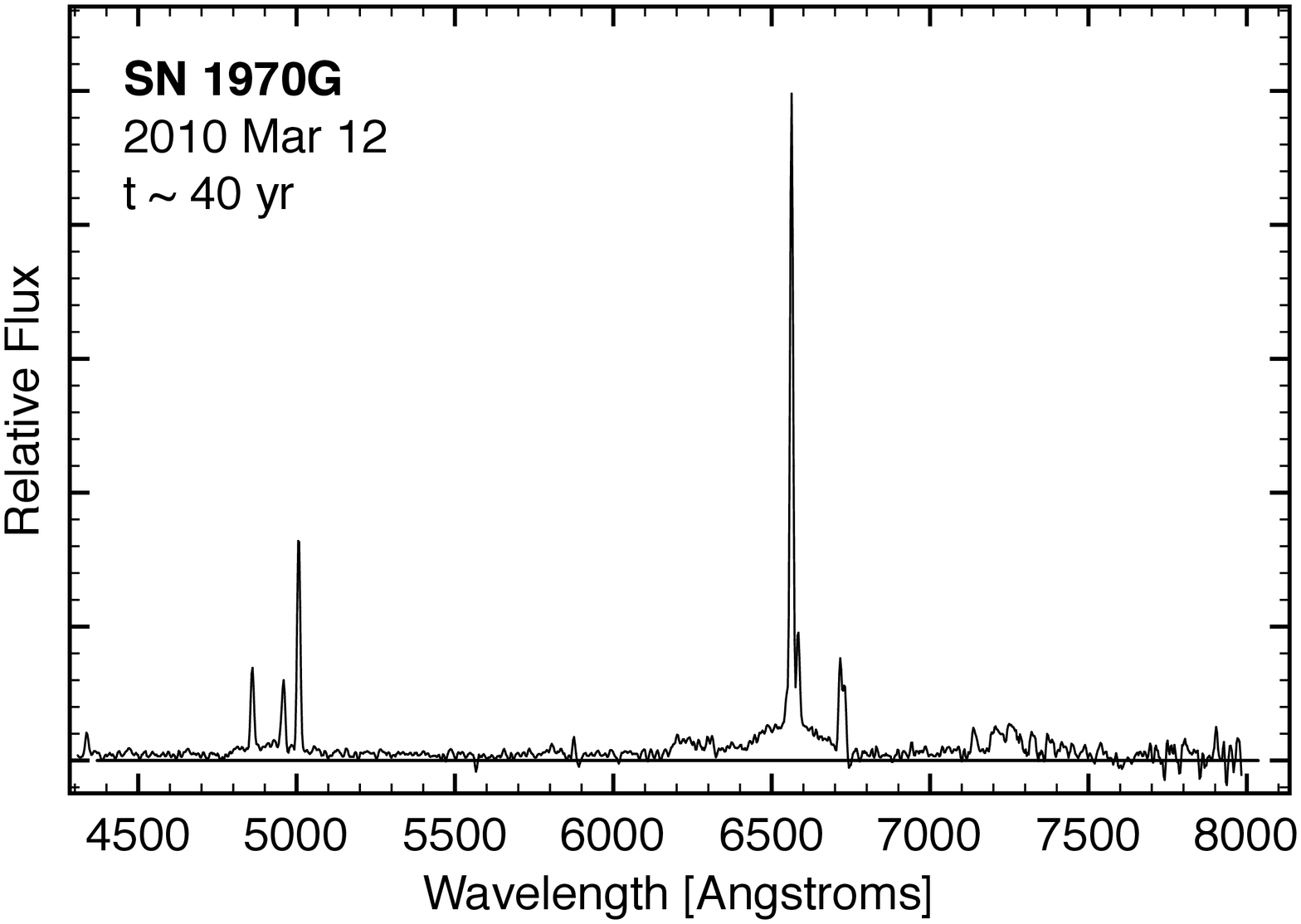}
\includegraphics[width=0.95\linewidth]{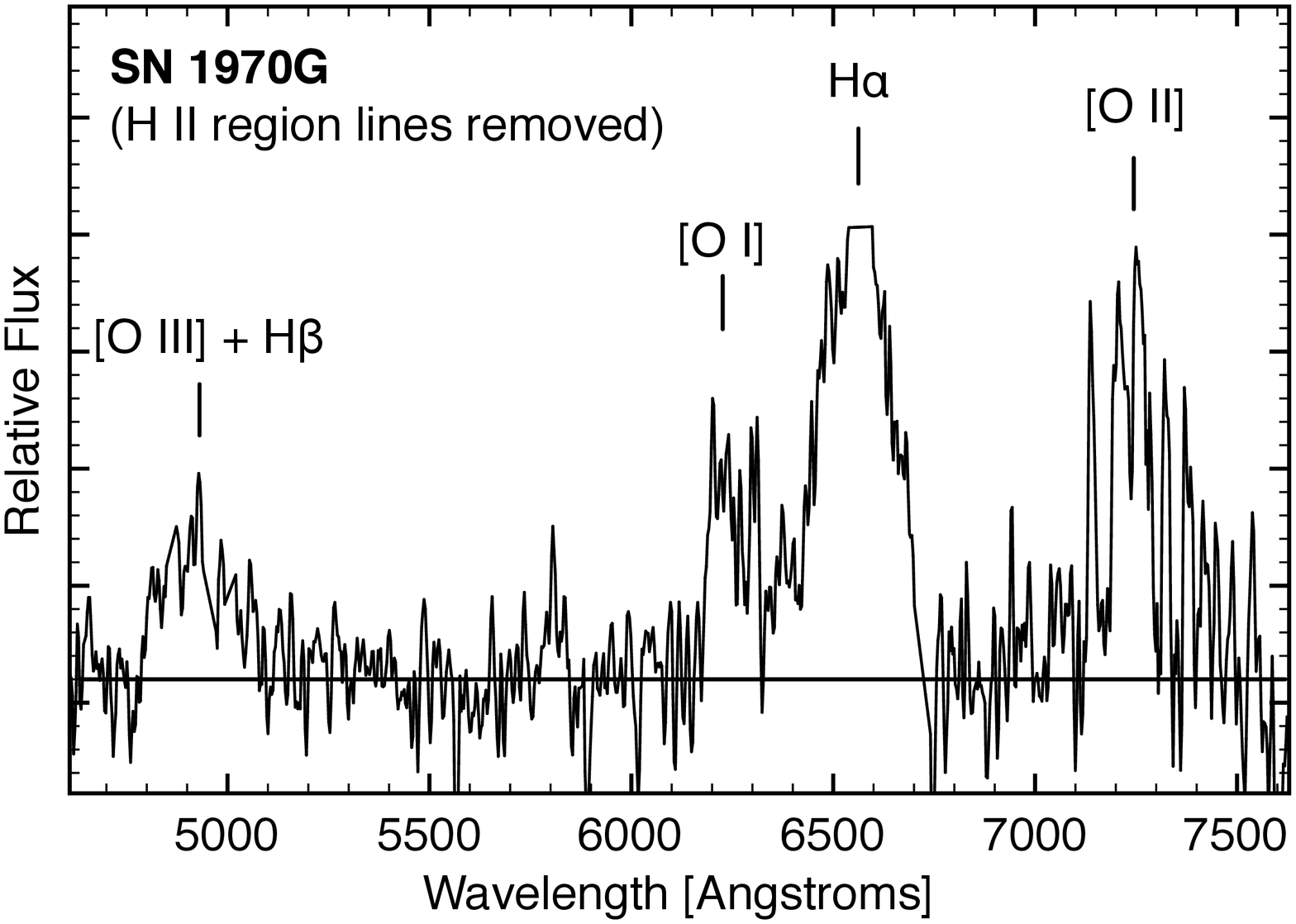}

\caption{{\it Top}: MDM optical spectrum of SN~1970G 40 yr post-outburst. The
  narrow emission lines dominating the spectrum are from a coincident H~II 
  region. {\it Bottom}: Enlarged spectrum with H~II region lines removed
  and line identifications marked. A blue continuum has been subtracted.}

\label{fig:sn70G_spec}
\end{figure}

SN 1970G was the first supernova to be detected in the radio and
remains the longest monitored radio supernova \citep{Stockdale01}.
Radio flux densities were relatively constant for 3 yr before
beginning to fade in 1974 \citep{Allen76,Brown78,Weiler86}. Radio
emission was re-detected in 1991 by \citet{Cowan91}, prompting optical
observations by \citet{Fesen9370G} who obtained spectra revealing
broad H$\alpha$ and [\ion{O}{1}] \dlambda 6300, 6364 emissions.

Two optical {\sl HST} images of the region around SN~1970G obtained in 1998
April are shown in Figure~\ref{fig:sn70G_image}.  To the south lies the large
\ion{H}{2} region NGC 5455. The F606W image shows three sources around the radio
coordinates $\alpha(2000)$=14$^h$03$^m$00$\fs$88
\ $\delta(2000)$=$+$54$\degr$14$\arcmin$33$\farcs$1 \citep{Stockdale01} centered
in the figure. Although a finding chart published in \citet{Fesen9370G} suggests
the brighter, eastern source as the SN, alignment of these images with a CFHT
MegaPipe image of the region (Group G002.308.620+60.034) having astrometric
accuracy $<0\farcs1$ favors the western source, and is the object marked in
Figure~\ref{fig:sn70G_image}.  The observed F606W/F656N flux ratio of this
object is in agreement with the spectrum of the SN which shows broad H$\alpha$
emission.

\begin{figure*}[!htp]
\centering
\includegraphics[width=0.95\linewidth]{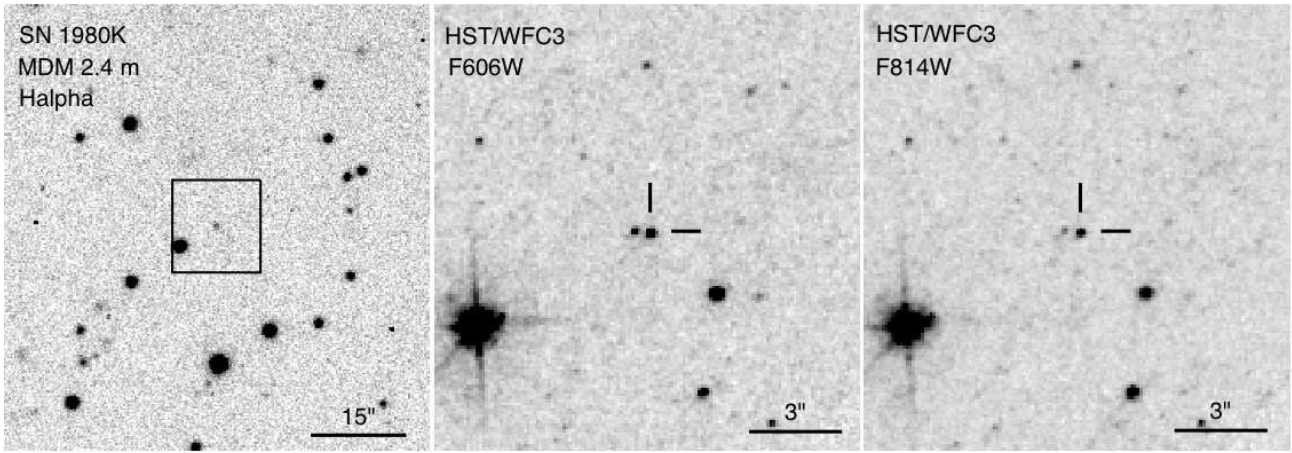} 

\caption{Optical images of SN~1980K. {\it Left}: H$\alpha$ image obtained in
1997 May with the 2.4 m Hiltner telescope showing the environment around
the SN. Boxed region demarcates enlarged portion shown in other panels.
{\it Middle and Right}: {\sl HST}/WFPC3 images obtained in 2010 March with SN
1980K marked.}

\label{fig:sn80K_images}
\end{figure*}

In the top panel of Figure~\ref{fig:sn70G_spec}, we show a 2010
optical spectrum of SN~1970G some 40 years after 
outburst. Wavelengths have been corrected for the host galaxy's
redshift of 251 \kms\ and an underlying blue continuum believed to be
associated with background starlight has been subtracted with a third
order Chebysev function. The spectrum is dominated by narrow
\ion{H}{2} region lines of [\ion{O}{3}], H$\alpha$, H$\beta$,
[\ion{N}{2}] \dlambda 6548, 6583, and [\ion{S}{2}] \dlambda 6716,
6731. Noise around 7300 \AA\ is due to poor night sky OH subtraction.

An enlarged portion of the spectrum with the \ion{H}{2} region lines manually
removed is shown in the lower panel of Figure~\ref{fig:sn70G_spec} in order to
better see the faint underlying emission originating from the supernova.
H$\alpha$ is the strongest emission line and exhibits a mildly asymmetric
profile shifted blueward. There is no clear change from the profile described in
\citet{Fesen9370G}. 

The estimated H$\alpha$ flux (minus background and narrow \ion{H}{2}
region emission) is $1.7 \pm 0.4 \times 10^{-15}$ erg s$^{-1}$
cm$^{-2}$, relatively unchanged from 1.8 $\times$ 10$^{-15}$ erg
s$^{-1}$ cm$^{-2}$ measured 18 years earlier \citep{Fesen9370G}.  The
width of H$\alpha$ is estimated to span 6410 \AA\ to 6690 \AA\
($-7000$ to $+5800$ \kms) with uncertainty perhaps greater than $\sim
500$ \kms\ due to confusion with the broad [\ion{O}{1}]
$\lambda\lambda$6300, 6364 and narrow [\ion{S}{2}] $\lambda\lambda$
6716, 6731 lines. The velocity width measured from data obtained in
1992 was reported to extend from $-5200$ \kms\ to $+5600$ \kms, but
this change is likely the result of improvement in S/N.

Broad emission spanning 4780 to 5090 \AA\ showing a strongly blueshifted,
gradually descending profile is detected.  We associate the emission primarily
with [\ion{O}{3}], but given the large implied blueshifted velocity ($< -10^4$
\kms) with respect to the 4959 \AA\ line, contribution from H$\beta$ is likely.

[\ion{O}{1}] emission shows the same gradually descending profile as
[\ion{O}{3}].  We measure a velocity extension from 6175 \AA\ ($-6000$
\kms\ with respect to 6300 \AA) to 6410 \AA\ ($+2200$ \kms\ with
respect to 6364 \AA) where it merges with the broad H$\alpha$.  Broad
emission centered around 7240 \AA\ could be associated with the
[\ion{Ca}{2}] \dlambda 7291, 7324 and/or [\ion{O}{2}] \dlambda 7319,
7330 lines \citep{Fesen9370G}.  Because [\ion{O}{2}] is known to be a
dominant line in intermediate-aged supernovae (see discussion in
\citealt{Fesen99}) and the velocity line profile best matches the
[\ion{O}{1}] distribution when centered with respect to 7325~\AA, we
identify [\ion{O}{2}] as the major contributor.

Faint emission centered around 5810 \AA\ having a full-width-at-zero-intensity
less than 100 \AA\ is also weakly detected. It was not detected in the
\citet{Fesen9370G} spectrum and may be associated with blueshifted \ion{He}{1}
and/or \ion{Na}{1} emission.

\begin{figure}[!htp]
\centering
\includegraphics[width=0.95\linewidth]{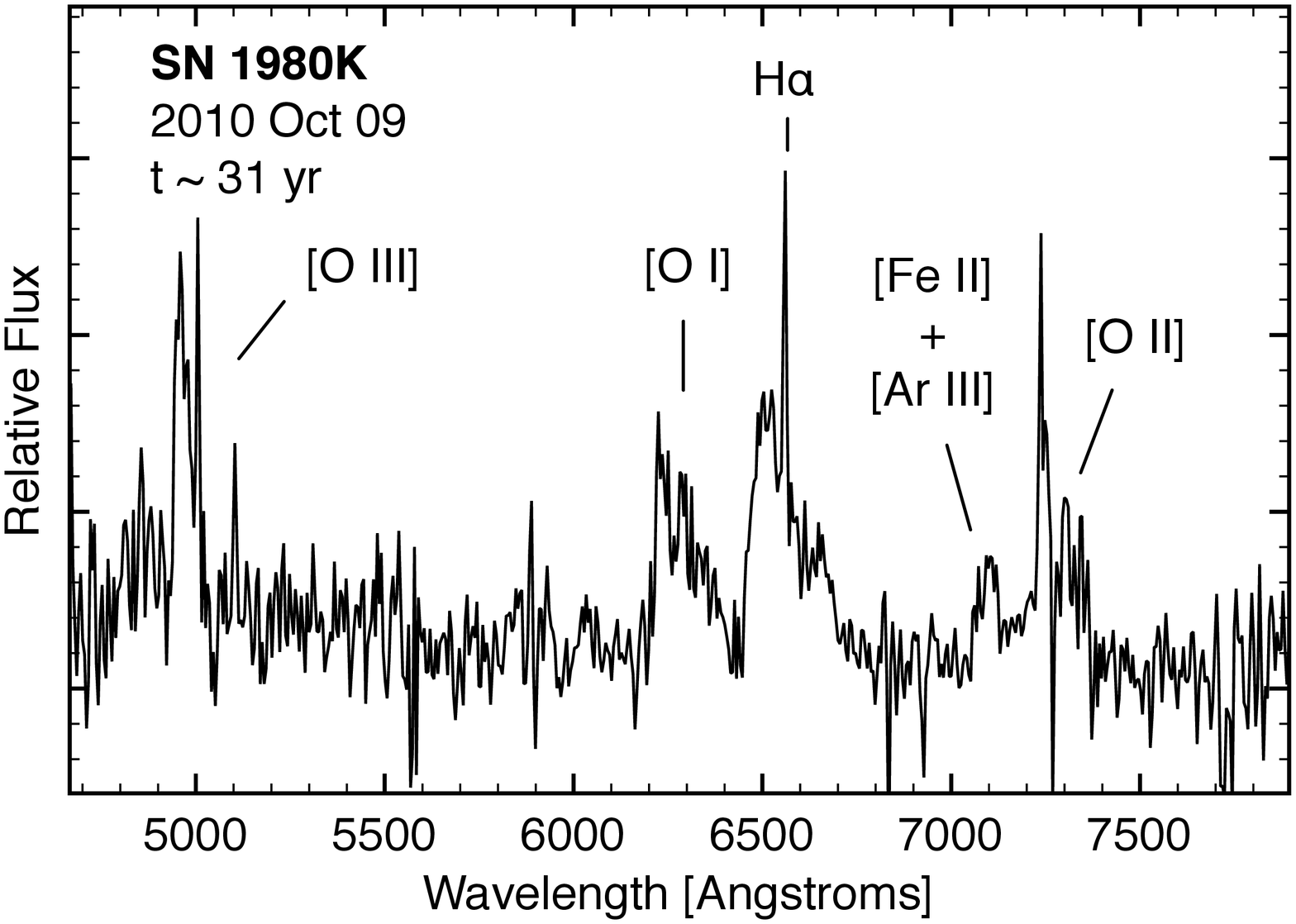}\\
\includegraphics[width=0.95\linewidth]{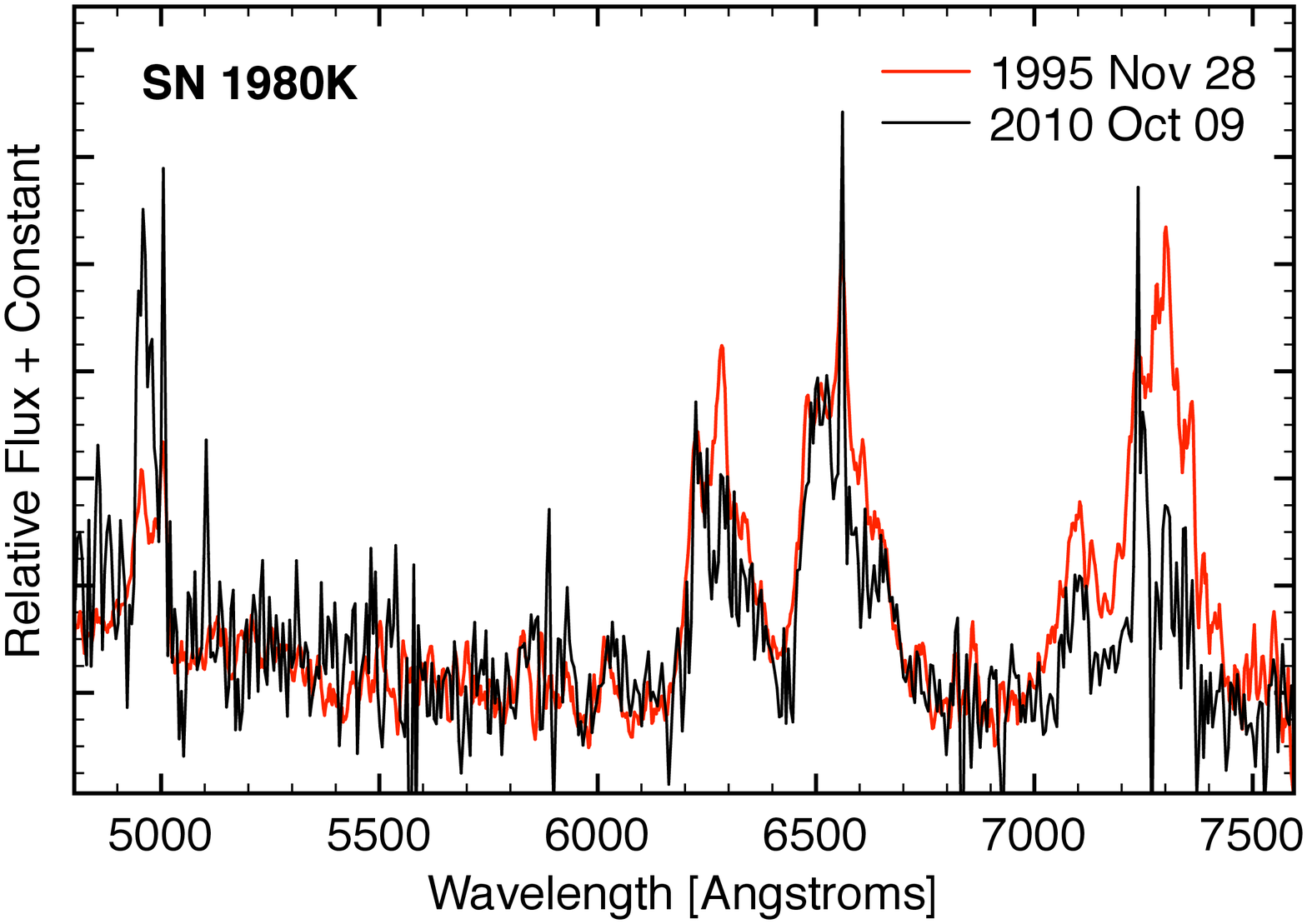}

\caption{{\it Top}: MDM optical spectrum of SN~1980K 31 yr post-outburst. {\it
    Bottom}: The 2010 spectrum of SN~1980K compared with a 1995 spectrum from
  \citet{Fesen99}.}

\label{fig:sn80K}
\end{figure}

\subsection{SN 1980K}

SN 1980K in NGC~6946 (D $\sim$ 5.9 Mpc; \citealt{Karachentsev00}) was discovered
by P.\ Wild on 1980 October 28 and reached a peak brightness of $V$ = 11.4 in
1980 November \citep{Buta82}. Photographic observations showing an almost linear
decline in brightness and spectra revealing broad H$\alpha$ with minor P Cygni
absorption classify it as a Type IIL \citep{Barbon82b}. Radio and X-ray
emissions were detected about a month after maximum
\citep{Canizares82,Weiler86}. Radio emissions were followed quite extensively
through 1996 \citep{Weiler92, Montes98}, and more limited X-ray observations
have been obtained through 2004 \citep{Schlegel94,Soria08}.

The SN declined steadily in the optical through 1982 \citep{Uomoto86} but was
detected in 1987 nearly seven years after maximum through narrow passband
imaging \citep{Fesen88}.  Follow-up low-dispersion optical spectra showed broad
H$\alpha$ and [\ion{O}{1}] emission along with weaker line
emission from [\ion{O}{3}], [\ion{Fe}{2}] $\lambda$7155, and
emission around 7300~\AA\ identified as [\ion{Ca}{2}] \dlambda 7291, 7324 and/or
[\ion{O}{2}] \dlambda 7319, 7330 \citep{Fesen90,Uomoto91,Leibundgut91}. Further
monitoring through 1997 indicated no major changes in the spectrum aside from
steadily declining H$\alpha$ emission \citep{Leibundgut93,Fesen94,Fesen99}.

High resolution {\sl HST} images obtained in 2010 March
(Fig.~\ref{fig:sn80K_images}) show the environment of SN 1980K in
detail never before seen. Two unresolved sources around the published
coordinates of the supernova $\alpha$(2000) = 20$^h$35$^m$30$\fs$07
$\delta$(2000) = +60$\degr$06$\arcmin$23$\farcs$8 \citep{vanDyk95} are
seen.  We compared the images to an unpublished ground-based H$\alpha$
image ($\lambda_{\rm c}$ = 6564~\AA; $\Delta\lambda$ = 30 \AA) of the
region obtained by R.~Fesen on 1997 May 08 using the Hiltner 2.4~m
telescope at MDM Observatory (also shown in
Fig.~\ref{fig:sn80K_images}) to confirm that the supernova was the
brighter, more westerly source. The other source, approximately
0$\farcs$55 away or 15 (d/5.9 Mpc) pc, has a rather steep color
difference between the F606W and F814W images suggestive of a luminous
blue star or very tight stellar association.

The top panel of Figure~\ref{fig:sn80K} shows our recent 2010 spectrum
of SN~1980K with identified emission lines marked. The small recession
velocity of 40 \kms\ of NGC~6946 has been corrected. A weak blue
continuum that trails off around around 5500 \AA\ may be due to light
contamination from the projected companion seen slightly east of the
SN in the HST/WFC3 images (Fig.~\ref{fig:sn80K_images}).

The H$\alpha$ and [\ion{O}{1}] lines exhibit broad, asymmetric
emission profiles strongly blueshifted with an emission peak around
$-2900$ \kms. Emission around 7300 \AA\ associated with [\ion{O}{2}]
\citep{Fesen99} is weakly detected.  Asymmetric emission spanning 4935
to 5010 \AA\ is identified as strongly blueshifted [\ion{O}{3}]. A
narrow unresolved emission peak around 6558 \AA, seen in earlier
spectra, is potentially due to a small \ion{H}{2} region near the SN
site or ionized wind material associated with the progenitor
\citep{Fesen90,Fesen95a}.  Another peak around 5003 \AA\ is also
observed; if associated with [\ion{O}{3}] $\lambda$5007, this is at
the same $\approx -200$ \kms\ blueshifted velocity of the narrow
H$\alpha$ line.  Emission around 7090 \AA\ is likely associated with
[\ion{Fe}{2}] $\lambda$7155 blueshifted $-2700$ \kms, but some
contribution from [\ion{Ar}{3}] $\lambda$7136 is possible.

\begin{figure}[htp!]
\centering
\includegraphics[width=0.95\linewidth]{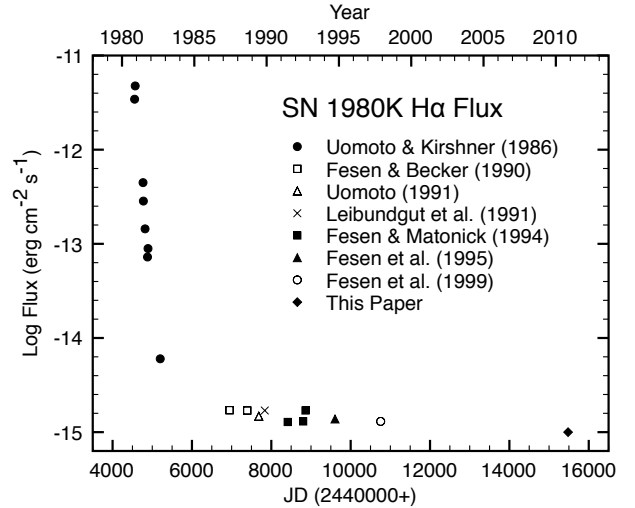}

\caption{Observed H$\alpha$ flux in SN 1980K over the period 1980--2010.}

\label{fig:sn80K_haflux}
\end{figure}

The bottom of Figure~\ref{fig:sn80K} compares our 2010 spectrum of
SN~1980K with Keck spectra obtained in 1995 \citep{Fesen99}.
[\ion{O}{1}] and H$\alpha$ show minor changes to their line profiles
and overall reduction in emission strength, while the flux ratio of
[\ion{O}{3}]/H$\alpha$ has increased. The overall [\ion{O}{2}] flux has
decreased, but the true extent of diminishment is difficult to gauge
firmly because of low S/N and poor night sky subtraction.

The H$\alpha$ line profile has a velocity width of $-5300$ to $+6000$ \kms.  The
blueshifted emission width is narrower than the 1995 measurement of $-5700$
\kms, and the redshifted width is slightly larger than the previous estimate of
$5500$ \kms.  The blueshifted velocity change does not follow the estimated rate
of $\approx-400$ \kms\ yr$^{-1}$ first noted by \citet{Fesen94} and later
confirmed by \citet{Fesen95a}. [\ion{O}{1}] has a velocity of $-4700$ \kms,
which is considerably smaller than the $-6000$ \kms\ reported in spectra from
1995. The substantial velocity change is likely associated with improved S/N.

We estimate a broad H$\alpha$ flux of $1.0 \pm 0.2 \times 10^{-15}$
erg s$^{-1}$ cm$^{-2}$, which is a small drop from $1.3 \pm 0.2 \times
10^{-15}$ erg s$^{-1}$ cm$^{-2}$ reported by \citet{Fesen99} 15 yr
earlier. In Figure~\ref{fig:sn80K_haflux}, a plot of all published
H$\alpha$ flux measurements shows an almost flat emission trend over
the past two decades.  Persistent H$\alpha$ emission with little to no
decline over several years is consistent the other SN Type IIL already
discussed, SN~1970G, and with the handful of additional Type IIL SNe
observed years post-outburst including SN 1979C
\citep{Fesen90,Fesen94,Fesen95a} and SN 1986E, whose brightness in $R$
at eight years of age was almost the same it was two years after
outburst \citep{Cappellaro95}. The origin of this emission has long
been attributed to reverse shock-heated hydrogen-rich ejecta (e.g.,
\citealt{Fesen99,Milisavljevic09}; see Section 5) although
\citet{Sugerman12} suggest SN 1980K's H$\alpha$ emission may have
substantial contribution from scattered light echoes.

\subsection{SN 1993J}

SN 1993J in M81 (D $\sim$ 3.6 Mpc; \citealt{Freedman94}) was discovered by
F.\ Garcia on 1993 March 28.9 \citep{Ripero93} and reached a maximum brightness
of $V$ = 10.8 mag \citep{Richmond94}. Early
spectra showed an almost featureless blue continuum with indications of broad
but weak H$\alpha$ and \ion{He}{1} $\lambda$5876 suggesting a Type II event
\citep{Filippenko93a,Garnavich93}.  However, the spectra quickly evolved to show
\ion{He}{1} lines associated with the Type Ib class
\citep{Filippenko93b}, thus acquiring the transitional
classification Type IIb \citep{Filippenko93c} anticipated by \citet{Woosley87}.

\begin{figure}[htp!]
\centering
\includegraphics[width=0.95\linewidth]{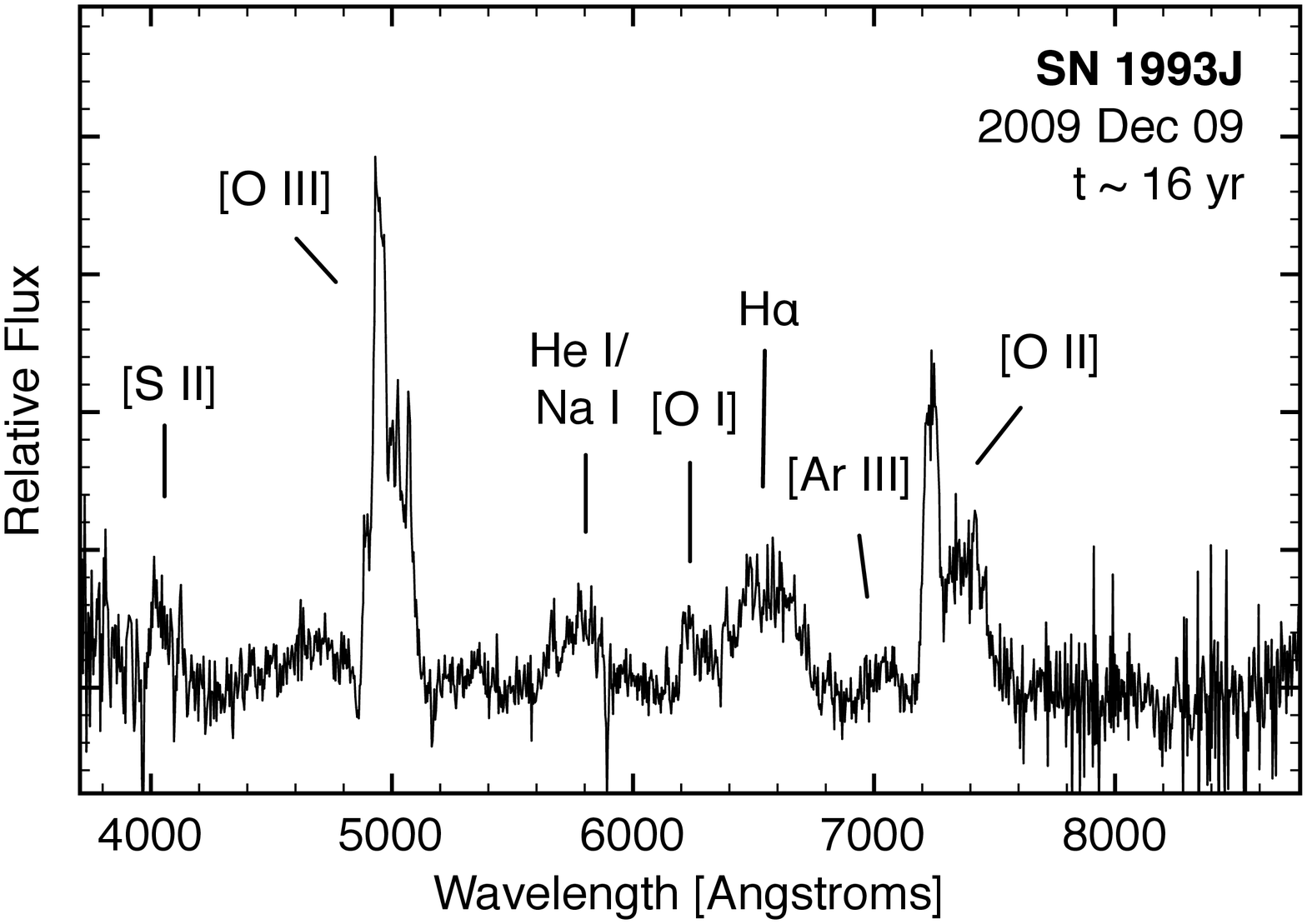}\\
\includegraphics[width=0.95\linewidth]{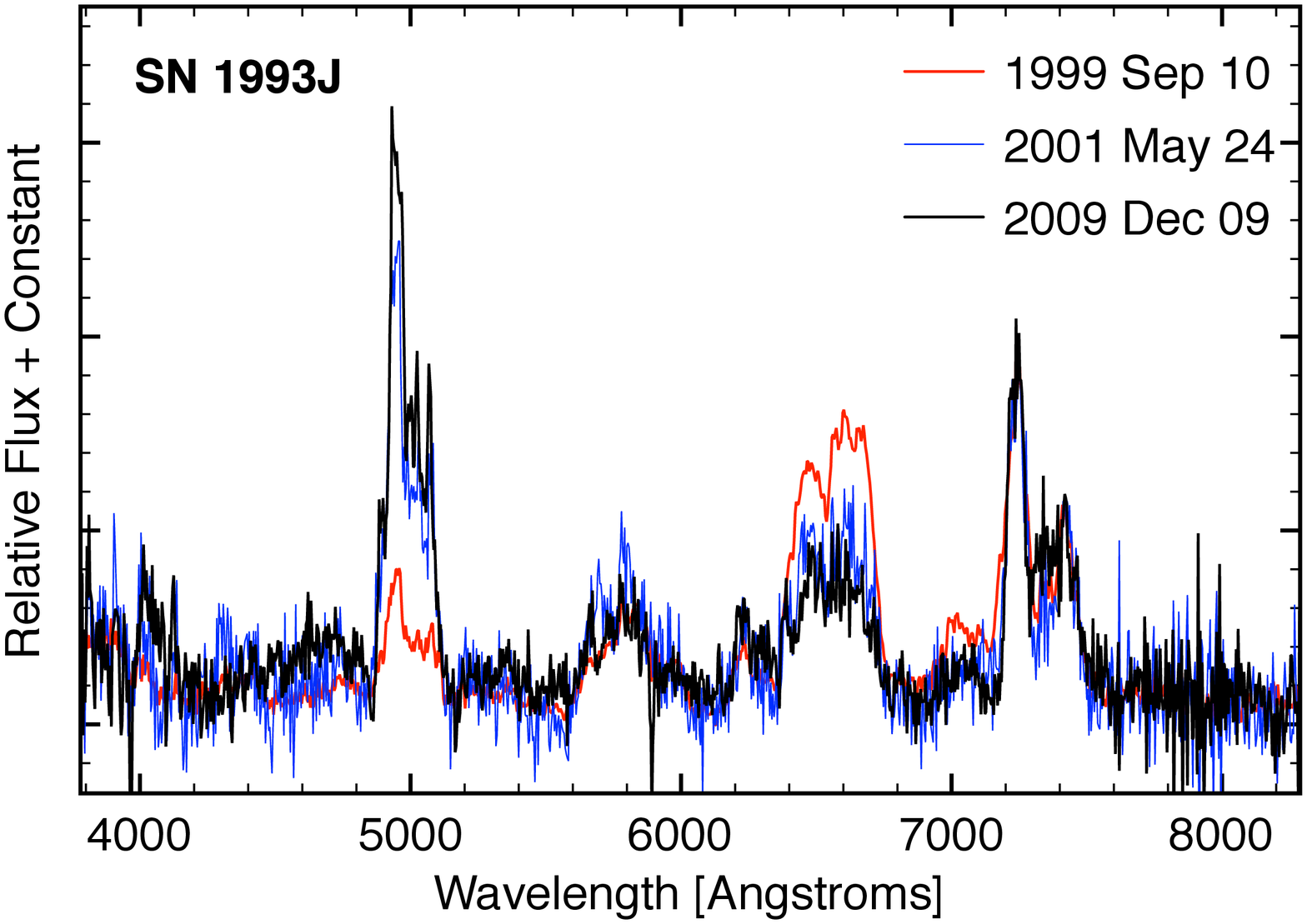}

\caption{{\it Top}: MMT optical spectrum of SN~1993J 16 yr post-outburst with
line identifications marked. {\it Bottom}: The 2009 spectrum of SN 1993J
(above) compared with a 1999 spectrum from \citet{Matheson00b,Matheson00a} and a
previously unpublished 2001 spectrum.}

\label{fig:sn93J}
\end{figure}

The relatively isolated location and nearby distance of SN 1993J has
enabled close monitoring of emission in the X-ray
\citep{Zimmerman94,Suzuki95,Chandra09} radio
\citep{Bartel94,vanDyk94,Bartel00,Weiler07}, and optical
\citep{Woosley94,Filippenko94,Matheson00b,Matheson00a,Filippenko03,Fransson05}.
Recently, photometric and spectroscopic observations a decade
post-outburst detected the signature of a potential massive binary
companion to the progenitor star \citep{Maund04,Maund09}.

In the top panel of Figure~\ref{fig:sn93J}, we present a 2009 December
spectrum of SN~1993J.  Identified emission lines have been marked,
wavelengths have been corrected for a recession velocity of $-140$
\kms\ \citep{Matheson00a}, and a blue continuum has been removed using
a third order Chebyshev function. In the lower panel of the figure,
our 2009 spectrum is plotted over and directly contrasted with one
obtained approximately 10 yr earlier by
\citet{Matheson00b,Matheson00a} and a previously unpublished MMT
spectrum obtained on 2001 May 24 kindly provided by T.\ Matheson and
M.\ Modjaz. All spectra have been normalized to the [O II] emission
lines to highlight the gradual increase in [\ion{O}{3}] emission with
respect to H$\alpha$.

\begin{figure}[htp!]
\centering
\includegraphics[width=0.75\linewidth]{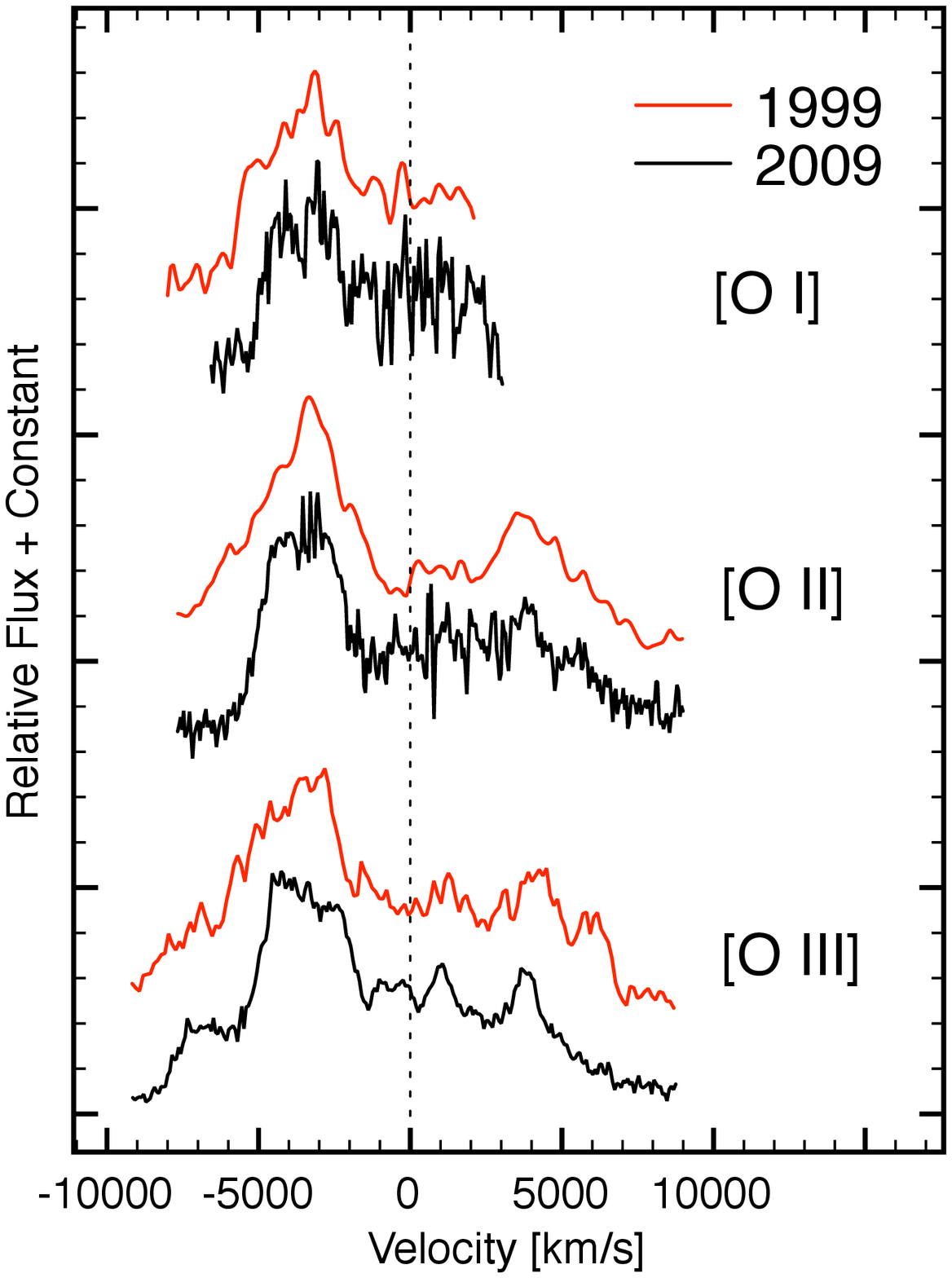}\\
\includegraphics[width=0.75\linewidth]{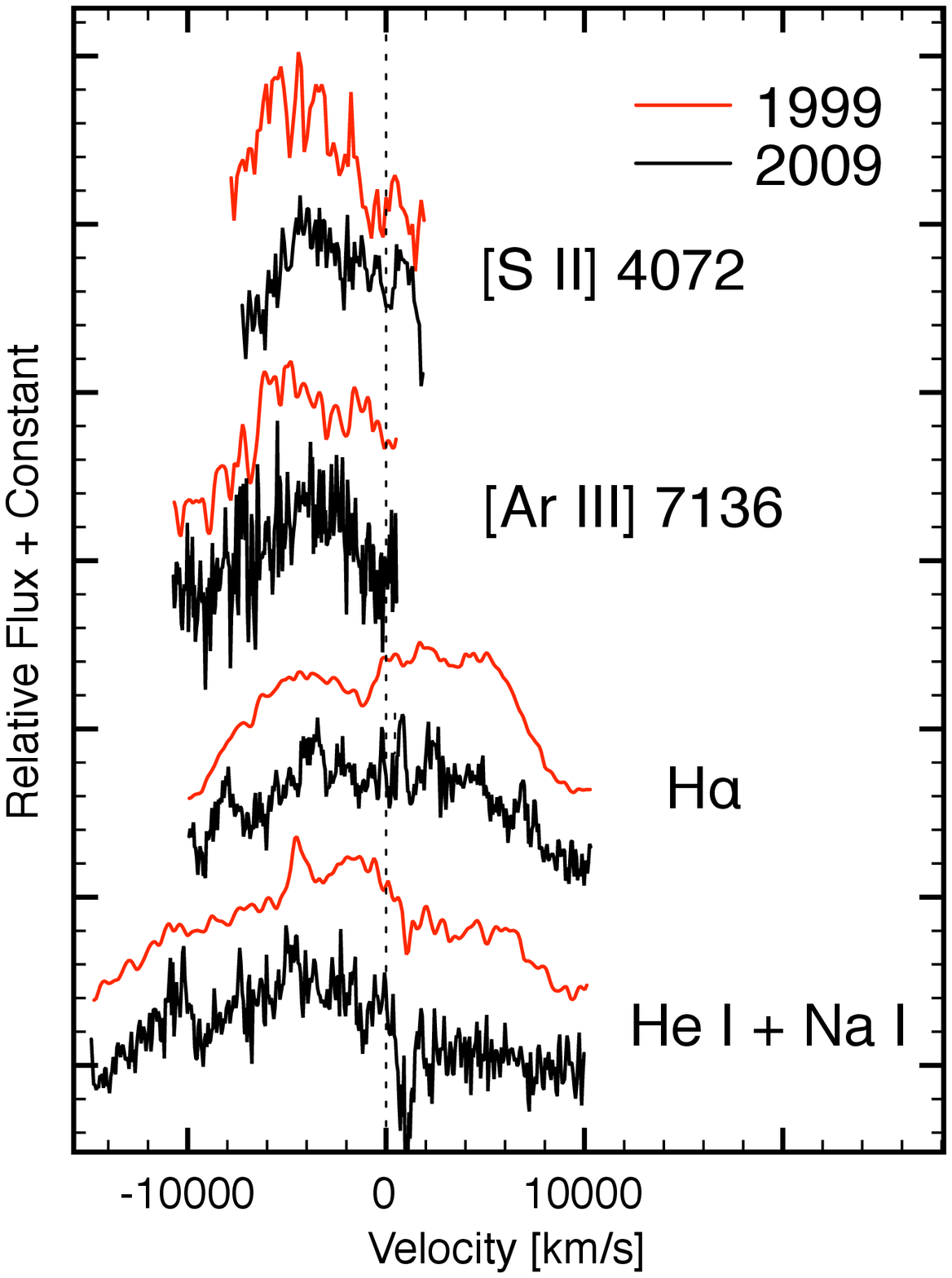}

\caption{Emission line profiles in the spectra of SN 1993J in 1999 and 2009.  In
  descending order, velocities are with respect to 6300, 7325, 5007, 4072, 7136,
  6563, and 5800 \AA. The 1999 spectrum is from \citet{Matheson00b,Matheson00a}.}

\label{fig:sn93J_lineevol}
\end{figure}

The general line profile shapes and small-scale features of the
strongest emission lines have not changed significantly (see
Fig.~\ref{fig:sn93J_lineevol}).  However, the ratio of
[\ion{O}{3}]/H$\alpha$ emission has increased by over an order of
magnitude, continuing a trend first noted by
\citet{Matheson00b}. Previous spectra showed a persistent H$\alpha$
line that, up until day 2452 after outburst, dominated the spectrum
with an unusual, box-like profile with velocities in excess of $\pm
9000$ \kms. However, [\ion{O}{3}] is now the dominant emission
feature. The profile maintains its pronounced asymmetry toward
blueshifted velocities with a major emission peak centered around
$-3500$ \kms. We estimate a total [\ion{O}{3}] \dlambda 4959, 5007
flux of $9.5 \pm 0.2 \times 10^{-15}$ erg s$^{-1}$ cm$^{-2}$, implying
a decline from the $1.6 \times 10^{-14}$ erg s$^{-1}$ cm$^{-2}$
measured in 1999.  The velocity span of $-5800$ to $+7000$ \kms\
has remained unchanged.

The [\ion{O}{2}] \dlambda 7319, 7330 lines follow the same profile distribution
as [\ion{O}{3}], as does [\ion{O}{1}] with the exception that at around
6365~\AA\ it begins to merge with H$\alpha$ (see
Fig.~\ref{fig:sn93J_lineevol}).  The [\ion{S}{2}] \dlambda 4069, 4076 lines
have become more pronounced, exhibiting a blueshifted profile sharply peaked at
$-3900$ \kms\ and gradually sloping down to zero velocity. The blend of
\ion{He}{1} and \ion{Na}{1} lines centered around 5800 \AA\ remains largely
unchanged, though evidence for an emission peak centered around 5670 \AA\ has
developed.

Emission around 7050 \AA\ has several possible sources of origin including
\ion{He}{1} $\lambda$7065, [\ion{Fe}{2}] $\lambda$7155, and [\ion{Ar}{3}]
$\lambda$7136. \citet{Matheson00b} attribute the feature to [\ion{Fe}{2}]
$\lambda$7155.  However, we found that centering the velocity distribution with respect to 7136
\AA\ provided a good fit with the [\ion{S}{2}] profile (see
Fig.~\ref{fig:sn93J_lineevol}).  Thus, assuming a common origin for Ar and S from the
Si-S-Ar-Ca interior layer of the progenitor star, [\ion{Ar}{3}]
$\lambda$7136 is likely the dominant contributor.

\begin{figure*}[htp!]
\centering
\includegraphics[width=0.70\linewidth]{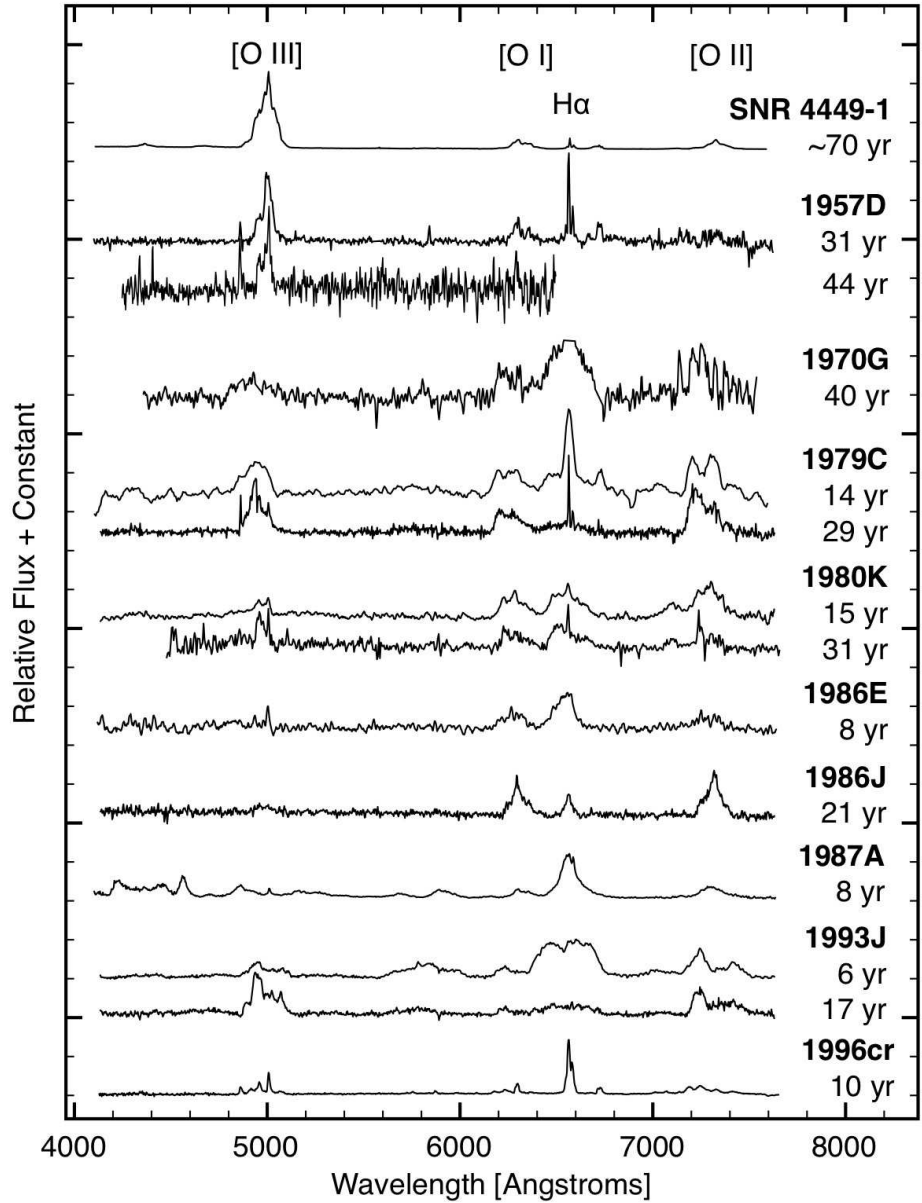}

\caption{Optical spectra of intermediate-aged core-collapse supernovae
exhibiting broad HWZI line widths $\ga 2000$ \kms.}

\label{fig:latetimespec}
\end{figure*}

\subsection{Additional Late-Time CCSN Spectra}
\label{sec:Latespec}

To help place the spectra of SN~1957D, 1970G, 1980K, and 1993J in context with
other CCSNe observed at similarly evolved epochs, we retrieved archival
late-time spectra for several CCSNe where broad (HWZI $\ga$ 2000 \kms) line
emission associated with metal-rich ejecta was observed at epochs $\ga$ 10 yr.
These objects include SN~1979C and SN~1986E (SN IIL), SN~1986J and SN~1996cr
(SN IIn), and SN~1987A (SN~IIpec).

Also included was the ultraluminous, O-rich supernova remnant in NGC~4449,
SNR~4449-1. SNR~4449-1 has no known classification, but properties of its
late-time spectra and stellar environment are in line with a stripped-envelope
progenitor \citep{Seaquist78,Kirshner80,Blair83,Blair84,Milisavljevic08a}.
Recent work has suggested an age of $\sim$70 yr, which is comparable to the
CCSNe studied here \citep{Bietenholz10b}.  Spectra of all ten SNe are shown in
Figure~\ref{fig:latetimespec} with their properties and references listed in
Table~\ref{tab:latespec}.

Not included in our sample were some Type IIn CCSNe where strong Balmer
emission associated with hydrogen-rich CSM dominates their spectra;
e.g., SN~1978K \citep{Ryder93,Schlegel99}, and SN~1988Z
\citep{Aretxaga99}. Also not included was the unique event SN~1961V
\citep{Stringfellow88} in light of its debated nature and limited
optical spectra \citep{Goodrich89,Filippenko95,Chu04,Kochanek11}.

\begin{deluxetable*}{ccccl}
\centering
\tabletypesize{\scriptsize}
\tablecaption{Details of Optical Spectra Presented in Figure~\ref{fig:latetimespec}}
\tablecolumns{5}
\tablewidth{0pc}
\tablehead{\colhead{SN}                  &
           \colhead{Type}                &
           \colhead{Host Galaxy}         &
           \colhead{Distance (Mpc)\tablenotemark{a}}      &
           \colhead{References}}
\startdata
SNR 4449-1 & \nodata  & NGC 4449 & 3.6*  & \citet{Bietenholz10b}\\
1957D & II     & M83        & 4.6  & \citet{Cappellaro95}; This paper\\
1970G & IIL    & M101       & 6.7  &  This paper\\
1979C & IIL    & M100       & 15.8 & \citet{Fesen99,Milisavljevic09}\\
1980K & IIL    & NGC 6946   & 5.9  & \citet{Fesen99}; This paper\\
1986E & IIL    & NGC 4302   & 24.9* & \citet{Cappellaro95}\tablenotemark{b} \\
1986J & IIn    & NGC 891    & 10.2* & \citet{Milisavljevic08b,Leibundgut91}\\
1987A & IIpec  & LMC        & 0.05 & \cite{Chugai97}\tablenotemark{c}\\
1993J & IIb    & M81        & 3.7  & \citet{Matheson00b,Matheson00a}; This paper\\
1996cr & IIn   & Circinus   & 4.2*  & \citet{Bauer08}
\enddata
\tablenotetext{a}{References to distance estimates provided in
  Sections \ref{sec:Intro} and \ref{sec:Results} unless marked with an asterisk (*) in which case they are mean distances
  reported by Nasa/IPAC Extragalactic Database
at http://ned.ipac.caltech.edu.}
\tablenotetext{b}{Data retrieved from SUSPECT at 
http://bruford.nhn.ou.edu/$\sim$suspect/index1.html.}
\tablenotetext{c}{Data retrieved from MAST at http://archive.stsci.edu/.}
\label{tab:latespec}
\end{deluxetable*}

\section{Discussion}
\label{sec:Discussion}

As can be seen from Figure~\ref{fig:latetimespec}, there is great
variety in the late-time optical spectra of intermediate-aged
CCSNe. Nonetheless, there exist some similarities between these
objects with respect to changes in emission line strengths and widths
over time. For example, the three SN~IIL objects SN 1970G, 1979C,
and 1980K all show broad H$\alpha$ ($\simeq$ 5000 km s$^{-1}$) with
the redshifted emission weaker than the blueshifted emission. These
spectra also show comparable evolutionary changes in the strength of
[\ion{O}{3}] relative to H$\alpha$.  On the other hand, quite
different late-time emissions are seen for objects such as SN 1986J
and 1987A. Below we describe several properties that were found to be
common to many of these spectra, and in Sections~\ref{sec:Models} and
\ref{sec:CasA} we interpret their physical origin in the context of a
SN--CSM interaction model and comparison to the young Galactic
supernova remnant Cassiopeia A.

\subsection{Changes in Relative Line Strength and Width Over Time}

In cases where multi-epoch spectra with good S/N are available, the
[\ion{O}{3}] \dlambda 4959, 5007 and H$\alpha$ lines change in
strength relative to the [\ion{O}{1}] \dlambda 6300, 6364 and
[\ion{O}{2}] \dlambda 7319, 7330 lines. Namely, the flux ratio
[\ion{O}{3}]/([\ion{O}{1}]+[\ion{O}{2}]) increases with time, and in
most cases H$\alpha$/([\ion{O}{1}]+[\ion{O}{2}]) decreases.  This is
clear in the spectra of three objects: SN 1980K between 1995 and 2010
(see Figure~\ref{fig:sn80K}), SN~1993J between 1999 and 2009 (see
Figure~\ref{fig:sn93J}), and in previously published spectra of
SN~1979C between 1993 and 2008 \citep{Milisavljevic09}. The fact that
emission from the oldest SNe in our sample, SN 1957D and SNR 4449-1,
are dominated by [\ion{O}{3}] emission with no appreciable H$\alpha$
is also consistent with the trend.

However, some SNe remain dominated by H$\alpha$ emission over all
observed epochs. For instance, SN~1970G is dominated by broad
H$\alpha$ emission as it has been since first observations by
\citet{Kirshner73}.  Although we detected [\ion{O}{3}] emission at
$t=40$ yr, its overall H$\alpha$ strength is far stronger than other
CCSNe of comparable ages.  In the same category, SN~1986E at 8 yr of
age continued to be H$\alpha$-dominated with insignificant
[\ion{O}{3}] emission. The unique objects SN~1986J and SN~1987A also
show weak or undetected [\ion{O}{3}] emission.

A handful of SN spectra in our sample show measurable changes in the
velocity widths of their emission line profiles. SN~1957D exhibited
narrowing in its [\ion{O}{3}] \dlambda 4959, 5007 lines of order 100
\kms\ yr$^{-1}$ (see Fig.~\ref{fig:sn57D_OIIIflux}), a rate similar to
changes in the velocity widths of forbidden oxygen emissions in SN
1979C between 1990 and 2008 \citep{Milisavljevic09}.  Narrowing of
order 25 \kms\ yr$^{-1}$ was also noted in SN~1979C's H$\alpha$
emission, but only affecting blueshifted velocities.  In some cases no
velocity changes were measured, (e.g., SN~1993J), while in other cases
changes in velocity width could not be measured confidently owing to
poor S/N and differences in detector sensitivity and/or spectral
resolution (e.g., SN~1970G).

Changes in velocity widths of emission line profiles are predicted to
be good tests between various mechanisms of late-time emission
\citep{ChevFran92,ChevFran94}. In circumstellar interaction scenarios
where the reverse shock penetrates into deeper layers of ejecta, the
velocity widths of emission lines are anticipated to
narrow. Alternatively, scenarios involving a pulsar wind nebula where
emission is powered by photoionization, line widths are anticipated to
broaden because of acceleration by the pulsar bubble. With one
exception, all SN studied here show either no discernible change in
their line widths or experienced narrowing of the order of $\sim$ 100
\kms\ yr$^{-1}$.

\begin{figure}[htp!]
\centering
\includegraphics[width=0.72\linewidth]{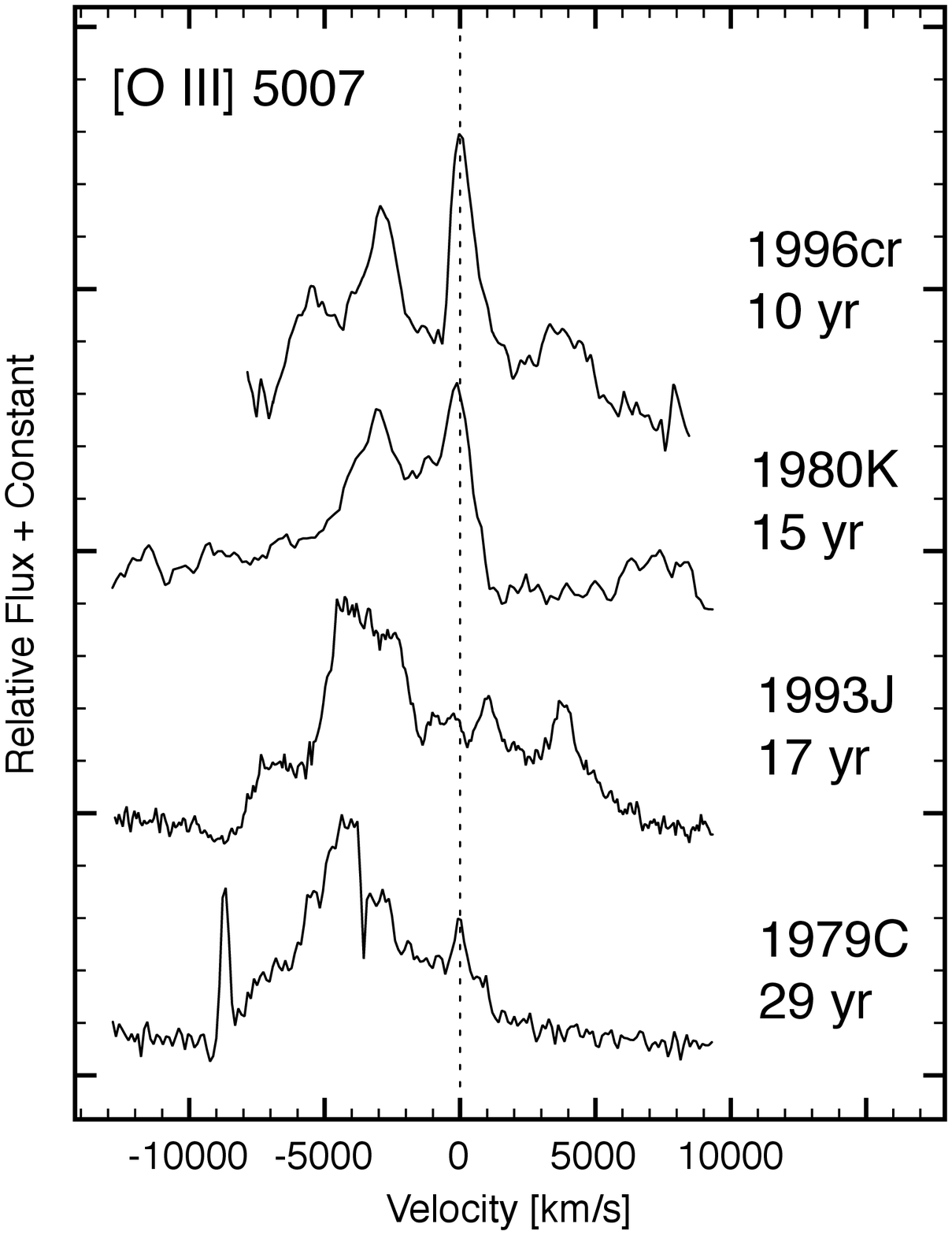}\\
\includegraphics[width=0.72\linewidth]{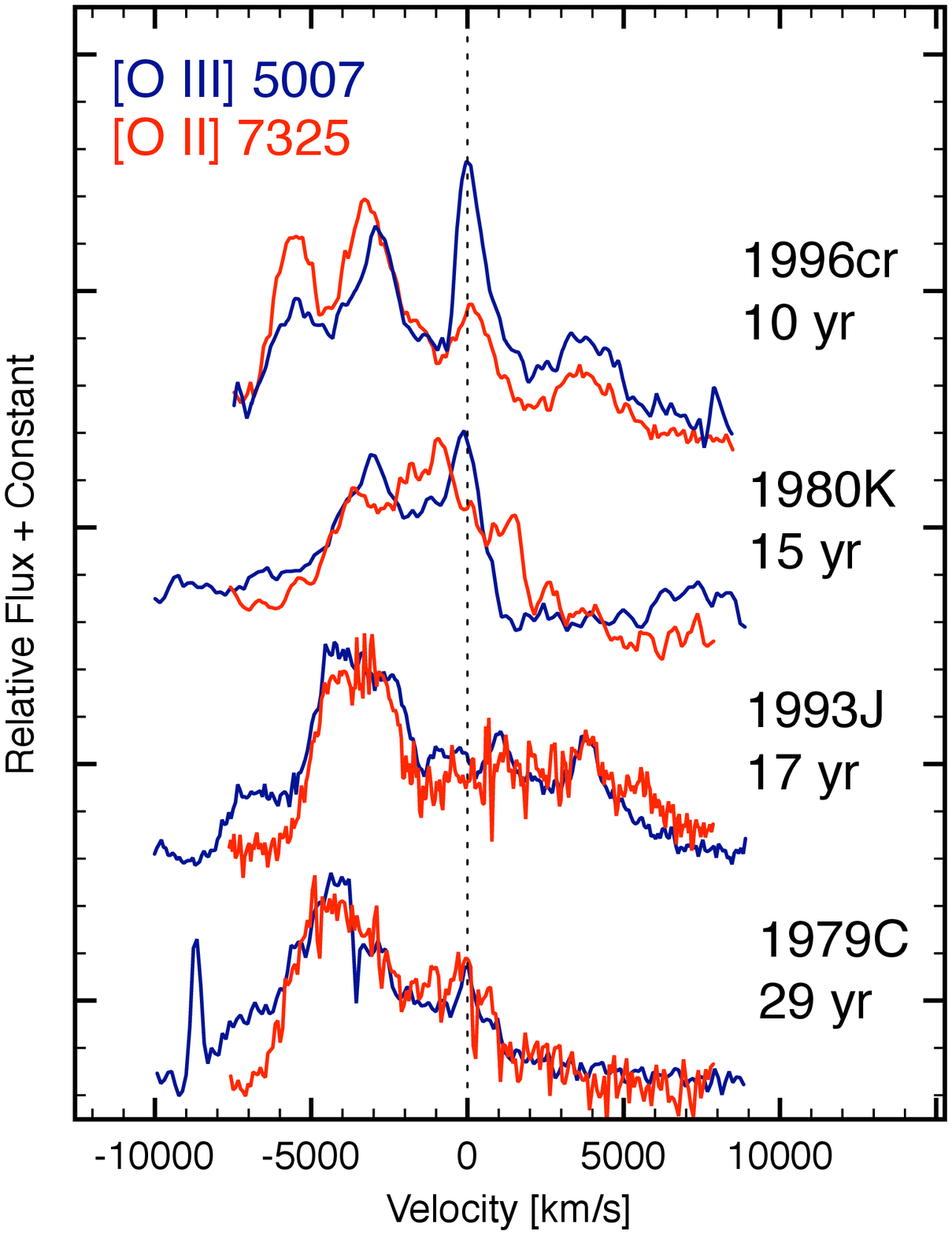}

\caption{Blueshifted emission line profiles from spectra presented in
Figure~\ref{fig:latetimespec}. {\it Top}: [O~III] \dlambda 4959, 5007 line
profiles. Velocities are with respect to 5007~\AA. {\it Bottom}: [O~III]
profiles compared to [O~II] \dlambda 7319, 7330 profiles from the same
epoch. Velocities are with respect to 7325 \AA.}

\label{fig:oxyprofiles}
\end{figure}

\subsection{Blueshifted Line Profiles}
\label{sec:Blueshifted}

Blueshifted line profiles are quite common in the late-time CCSN
spectra. Emission asymmetry is strongest in the forbidden oxygen lines,
particularly [\ion{O}{3}], but is also observed in H$\alpha$ line profiles where
it is generally less pronounced. The oxygen line profiles also show conspicuous
emission line substructure.

The presence of blueshifts in many late-time CCSN spectra is
illustrated in the top panel of Figure~\ref{fig:oxyprofiles}, where we
show line profiles of [\ion{O}{3}] emission for SN 1979C, 1980K,
1993J, and 1996cr. These SNe exhibit noticeably asymmetric profiles
dominated by blueshifted emission with velocities in excess of $-3000$
\kms. SN 1979C and 1980K, showing very little redshifted emission, are
extreme examples of this phenomenon. In the case of SN~1996cr,
evidence for two blueshifted peaks can be seen. The blueshifted
asymmetry and line substructure is also present in [\ion{O}{2}] line
emission, which is shown in the bottom panel of
Figure~\ref{fig:oxyprofiles}.

These asymmetric oxygen emission line profiles suggest that internal absorption
is obscuring emission from the SN's receding rear hemisphere. One candidate for
this type of absorption at late epochs is dust formation in metal-rich ejecta
much like that seen in SN~1987A (see \citealt{McCray93} and references therein).
Beginning around 450 days after optical maximum, SN~1987A exhibited an increase
in infrared emission accompanied shortly thereafter with a noticeable blueshift
in the [\ion{Mg}{1}], [\ion{O}{1}], and [\ion{C}{1}] emission lines.  The
blueshift was attributed to an attenuation of emission from the receding side of
the ejecta by dust grains formed within the ejecta, and the infrared excess
associated with the subsequent re-radiation by the dust grains.

Absorption due to dust in SN ejecta has also been suggested to explain the late-time
asymmetric UV line profiles observed in SN 1979C \citep{Fesen99}. Double-peaked
profiles with the blueward peak substantially stronger than the red were seen in
\ion{C}{2}] \dlambda 2324, 2325, [\ion{O}{2}] \dlambda 2470, 2470, and
  \ion{Mg}{2} \dlambda 2796, 2803 emissions. A handful of additional cases where
  dust formation in ejecta has been proposed include SN 1990I
  \citep{Elmhamdi04}, 1999em \citep{Elmhamdi03}, 2003gd
  \citep{Sugarman06,Meikle07}, 2005af \citep{Kotak06}, 2007it \citep{Andrews11},
  and 2007od where it may have had a clumpy structure
  \citep{Inserra11,Andrews10}.

Alternatively, absorption by dust may take place in the cool dense shell (CDS)
in the shock region \citep{Denault03}.  This is a likely place for dust
formation because of its high density and low temperature
\citep{Fransson84,ChevFran94}.  Assuming the CDS is formed behind the reverse
shock (in the Lagrangian reference frame), both the front- and back-side line
emission will be absorbed by the dust, though hydrodynamic mixing giving rise to
a clumpy CDS can allow some emission to penetrate \citep{Chevalier95}.
Formation of dust in a CDS has been reported in at least three supernovae within
months of outburst including 1998S \citep{Gerardy00}, 2005ip
\citep{Fox09,Smith09}, and 2006jc \citep{Smith08}. It was also reported in SN
2004et, which exhibited a wide $\pm 8500$ \kms\ box-shaped H$\alpha$ emission
about two years past outburst with a blueshifted profile resembling many of the
oxygen profiles presented here (\citealt{Kotak09}; see Fig.~4 of that paper).
 
However, it is important to note that not all supernovae in our sample show
large and pronounced blueshifted oxygen or hydrogen line emission.  SN~1957D,
1986J, 1987A, and 1986E only exhibit relatively small velocity blueshifts of
approximately $-500$ \kms.  Another exception is the ultraluminous oxygen-rich
supernova remnant SNR 4449-1. Unlike the other supernovae of this group,
pronounced asymmetry is seen in SNR~4449-1's [\ion{S}{2}] \dlambda 6716, 6731
and [\ion{Ar}{3}] $\lambda$7136 emission profiles which exhibit blueshifted
velocity distributions spanning $-2500 < V_{\rm exp} < 500$
\kms\ \citep{Milisavljevic08a,Bietenholz10b}.

\section{Models of SN--CSM Interaction}
\label{sec:Models}

The optical emissions from the supernovae discussed here are likely to
be from circumstellar interaction, except for SN 1987A where
radioactivity is a plausible power source at an age of 8 years
\citep{Chugai97,Kozma98a,Kozma98b,Larsson11}.  Thus, the changes in
relative line strengths and widths of the O and H lines noted in
Section~\ref{sec:Results} are useful tests of SN--CSM interaction
models.

If the supernova and circumstellar gas is evenly distributed, the interaction is
expected to lead to brighter optical emission from the reverse shock region as
opposed to the forward shock because of the higher density there
\citep{ChevFran94,Chevalier03}.  The forward shock generally is nonradiative,
but radiative shocks can occur in this region if the circumstellar gas is very
clumpy.  This is the likely source of the narrow lines in Type IIn supernovae
\citep{Chugai94}.

\begin{deluxetable}{cccccc}
%\hline
\centering
\tablecaption{Shocked Ejecta Masses at the Epochs of Spectra}
\tablecolumns{6}
\tablewidth{0pt}
\tablehead{\colhead{SN}                 & 
           \colhead{Type}               &
           \colhead{$\dot M_{-5}/v_{w1}$}&
           \colhead{Age}                &
           \colhead{$M_{ej}/V_4$}        &
           \colhead{H line}\\
           \colhead{}                   &
           \colhead{}                   &
           \colhead{}                   &
           \colhead{(yr)}               &
           \colhead{($M_{\odot}$)}       &
           \colhead{}}   
%\hline
\startdata
1993J &  IIb &  2.4 &  6  & 0.1 &  present   \\
1993J &  IIb &  2.4 &  17  & 0.3 &  weak   \\
%% Cas A &  IIb &  2 &  320  &  5 &  weak   \\
1970G &  IIL &  6.8 &  40  & 2.2 &  present   \\
1979C &  IIL &  11 &  14  & 1.3 &  present   \\
1979C &  IIL &  11 &  29  & 2.6 &  weak   \\
1980K &  IIL &  1.3 &  31  & 0.3 &  present   \\
1986E &  IIL &  3 &  8  & 0.2 &  present   \\
\enddata

%\hline
\label{tab:Shocked}
\end{deluxetable}

Supernovae of Type IIL and IIb show broader optical line emission and are
plausibly described by the scenario of reverse and forward shocks.  Estimates of
the mass loss density for these events can be obtained from radio observations
\citep{Weiler02,Montes97} and are shown in Table~\ref{tab:Shocked}, where $\dot
M_{-5}$ is the progenitor mass loss rate in units of $10^{-5}~M_{\odot} ~{\rm
  yr^{-1}}$ and $v_{w1}$ is the wind velocity in units of 10 km s$^{-1}$.
Several assumptions are made in obtaining these estimates; in particular, the preshock
wind temperature is assumed to be $2\times 10^4$ K and the shock velocity is
$10^4$ km s$^{-1}$.  The value for SN 1986E is from \cite{Montes97} and is
scaled to the lower limit values of \cite{Weiler02} on SN 1979C and SN 1980K.
Overall, the wind densities are at the high end of wind densities estimated for
red supergiant stars.

In the initial phase of interaction, the gas heated at the reverse
shock radiatively cools and there is optical emission from both the
cooling shock wave and preshock ejecta that have been radiatively
heated \citep{ChevFran94,Chevalier03}.  During this phase, the shock
luminosity drops approximately as $v_{sh}^3$, where $v_{sh}$ is the
forward shock velocity.  The transition to a noncooling phase occurs
at
\begin{eqnarray}
t_{cr}= && 0.0017(n-3)(n-4)(n-2)^{3.34} \nonumber \\
&& \times\left(V_{ej}\over 10^4{\rm~km~s^{-1}}\right)^{-5.34}
\left(\frac{\dot M_{-5}}{v_{w1}}\right){\rm~day},
\label{cool}
\end{eqnarray}
where $n$ is the power law exponent of the ejecta density profile and $V_{ej}$
is the highest velocity in the freely expanding ejecta.  It can be seen that the
result is very sensitive to $n$ (expected to be in the range $7-12$) and the
velocity.  By this estimate, the cases considered here are expected to be in the
nonradiative phase, although \cite{Nymark09} model SN 1993J as being in the
radiative phase at an age of just 8 years.  

Another factor is the composition of matter at the reverse shock.
Equation (\ref{cool}) assumes solar abundances and an enhancement of
heavy elements prolongs the radiative phase.  The optical emission is
expected to be mainly from the pre-reverse shock ejecta.  Some
emission from the cool dense shell built up during the radiative phase
is possible, but that gas is expected to disperse with time.

This model allows an estimate of the circumstellar gas that has been swept up by
the forward shock front
\begin{equation}
M_{cs}=\frac{\dot M}{v_w} Vt=0.01 \dot M_{-5}v_{w1}^{-1}V_4t_{yr}~M_{\odot},
\end{equation}
where $V$ is the average velocity of the forward shock and $V_4$ is in units of
$10^4$ km s$^{-1}$.  The mass of shocked ejecta, $M_{ej}$, is simply related to
$M_{cs}$ if the supernova density profile is a power law with index $n$
\citep{Chevalier82b}.  For $n=7$, we have $M_{ej}/M_{cs}=0.82$, which is the
value used in Table~\ref{tab:Shocked}.  This value of $n$ is smaller than the
high value that might be expected at the outer part of the supernova because the
reverse shock has moved in to the supernova at these late times.  The mass ratio
is a factor $\sim2$ smaller for $n=6$, and the same factor larger for $n=9$.

Within these uncertainties, the results in Table~\ref{tab:Shocked} are
consistent with the view that the decline of H emission in these
supernovae, where it occurs, is due to all of the H envelope having
passed through the reverse shock wave.  In the case of SN 1993J,
models of the early supernova indicate a H envelope mass of
$\sim0.2~M_{\odot}$ \citep{Woosley94}. This mass would have been
shocked by the time of the observation at $t=17$ yr, so there is no
longer H present at the reverse shock.  Type IIL supernovae are
believed to have higher H envelope masses than Type IIb events. For
instance, \cite{Blinnikov93} find that H envelopes with mass $\sim
1-2~M_{\odot}$ are present.  Thus, the disappearance of the broad H
emission in SN 1979C is consistent within this framework.  The
observations of the H line compared to O lines are generally in accord
with expectations, and there is the prediction that the H line in SN
1970G should fade in the not-too-distant future.

Another expectation of the circumstellar interaction scenario is that the
interaction leads to the deceleration of the gas near the reverse shock front.
For the power law density profile, we have a radial expansion $R\propto t^m$,
where $m=(n-3)/(n-2)$ \citep{Chevalier82b}.  The highest velocity at the reverse
shock evolves as
\begin{eqnarray}
\frac{dV_{ej}}{dt} = && \nonumber \\
-(1-m)\frac{V_{ej}}{t} = && -100\left(\frac{1-m}{0.2}\right) \left(V_{ej}\over 10^4{\rm~km~s^{-1}}\right) \nonumber \\
&& \times\left(\frac{t}{20{\rm~yr}}\right)^{-1} {\rm~km~s^{-1}~yr^{-1}},
\end{eqnarray}
where $n=7$ is used for the reference value of $m$.
The expected rate of line narrowing is close to that observed.

\begin{figure}[htp!]
\centering
\includegraphics[width=\linewidth]{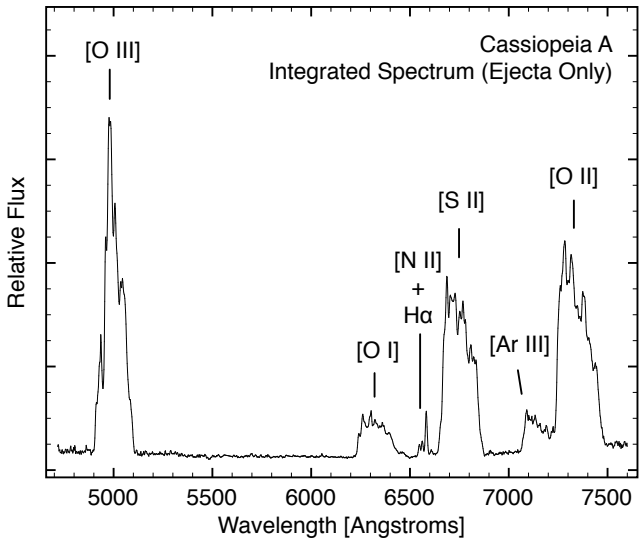}

\caption{Spatially integrated spectrum of the main shell of the Cassiopeia A
supernova remnant.} 

\label{fig:casa_intspec} 
\end{figure}

\section{CCSNe and SNRs: Comparisons to Cas A}
\label{sec:CasA}

In an effort to understand the nature of blueshifted and multiple peak emission
line profiles and relate their features with kinematic properties of the
optically emitting ejecta, a spatially integrated spectrum of the young Galactic
supernova remnant Cassiopeia A (Cas~A) was constructed.  Cas~A is the prototype
for the class of young, oxygen-rich SNRs and provides a clear look at the
explosion dynamics of a CCSN \citep{vandenBergh88,Fesen01}. Recent detection of
optical echoes of the supernova outburst indicate it was a Type IIb event,
probably from a red supergiant progenitor with mass $10-30$ M$_{\odot}$ that may
have lost much of its hydrogen envelope to a binary interaction
\citep{Krause08,Rest11,Fabian80,Vink96,Young06}. A nearby distance of 3.4 kpc
has permitted detailed study of Cas~A's composition and distribution of
supernova ejecta on fine scales \citep{Reed95,Fesen01WFPC2}, and its estimated
current age of $\approx$330 yr places it at a stage of evolution not that
different from the intermediate-aged CCSNe of our sample
\citep{Thorstensen01,Fesen06a}.

Our integrated spectrum of the Cas~A SNR is presented in
Figure~\ref{fig:casa_intspec}.  The spectrum was extracted from 80 long slit
spectra spaced $3''$ apart across the entire main shell (approximately
$4\arcmin$ in diameter) covering the wavelength region $4500-8000$
\AA\ (resolution 7 \AA) from an investigation of the remnant's three-dimensional
kinematic structure \citep{Milisavljevic12}.

This spectrum has not been corrected for reddening which can vary $4 \la
\rm{A}_{V} \la 8$ \citep{Hurford96}, so [\ion{O}{3}] emission is intrinsically
much more dominant than the presented spectrum indicates. These spectra also
largely omit emission from the remnant's slow-moving circumstellar material
called quasi-stationary flocculi (QSFs). Additionally, the spectrum does
not include emission from the faint NE jet or other outer ejecta which
contribute relatively insignificant optical flux.

\begin{figure}[htp!]
\centering
\includegraphics[width=0.75\linewidth]{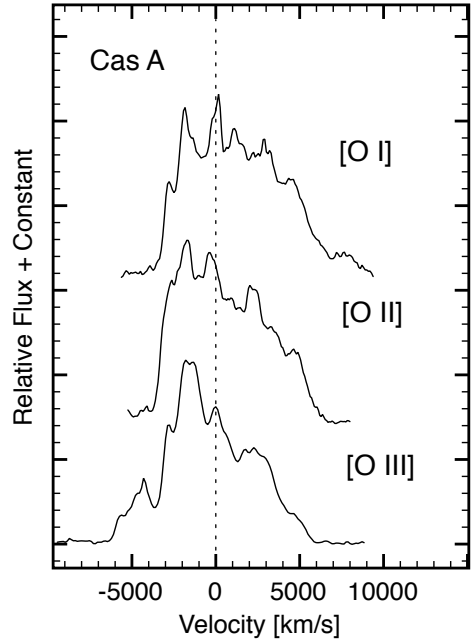}

\caption{[O~I] \dlambda 6300, 6364, [O~II] \dlambda 7319, 7330, and [O~III]
\dlambda 4959, 5007 line profiles from the integrated spectrum of Cassiopeia
A.}

\label{fig:casa_oxygen}
\end{figure}

\subsection{Cas A's Asymmetric Emission Profiles}

The Cas~A spectrum (Fig.~\ref{fig:casa_intspec}) shows pronounced blueshifted
emission with conspicuous line substructure in [\ion{O}{1}], [\ion{O}{2}],
[\ion{O}{3}], [\ion{S}{2}], and [\ion{Ar}{3}]. These line profiles are all
asymmetric, with bulk blueshifts peaked around $-1700$ \kms\ and a blue to red
asymmetry that increases somewhat with higher ionization levels. The velocity
line profiles of the forbidden oxygen lines are shown in
Figure~\ref{fig:casa_oxygen}.  The [\ion{S}{2}] \dlambda 6716, 6731 and
[\ion{Ar}{3}] $\lambda$7136 lines, not shown in Figure~\ref{fig:casa_oxygen},
have comparable blueshifts of around $-1700$ \kms.

The Cas~A spectrum appears similar to many of the intermediate-aged
CCSNe exhibiting strong [\ion{O}{3}] emission.  Particularly
well-matched with Cas~A are SN~1979C, SN~1993J, SNR 4449-1, and to
some degree SN~1980K (see Fig. \ref{fig:oxyprofiles}), which show
strong blueshifted oxygen emissions and/or conspicuous line
substructure. 

In Figure~\ref{fig:casa_comparison}, the [\ion{O}{3}] emission line
profile of the integrated Cas~A spectrum is plotted along with the
profiles of SN~1993J and SNR~4449-1 to highlight their similar
spectral features. It is interesting to note that Cas~A and SN 1993J
were both Type IIb SNe at photospheric stages, and here we see that
they share late-time spectral features as well.

Extensive multi-wavelength studies of Cas~A allow strong ties to be made between
the emission asymmetry and absorption due to dust in the ejecta. Cas~A is the
only Galactic supernova remnant that exhibits clear evidence of dust formed in
its ejecta \citep{Lagage96,Arendt99,Dwek04,Krause04}. Estimates of the total
mass remain controversial, ranging over several orders of magnitude from less
than $3 \times 10^{-3}$ M$_{\odot}$ to over 4 M$_{\odot}$. However, most
recently {\sl Spitzer} infrared observations showed close overlap between the
ejecta and dust maps of the remnant indicating freshly formed dust in the ejecta
having a total dust mass in the range of 0.020 - 0.054 M$_{\odot}$
\citep{Rho08}.

\subsection{Ejecta Rings: Are They Common?}

The three-dimensional kinematic properties of Cas~A's ejecta have been
studied in considerable detail
\citep{Lawrence95,Reed95,DeLaney10,Milisavljevic12}. In general,
Cas~A's optical emission takes the form of an approximately spherical
shell made up of several large ring-like structures composed of O- and
S-rich material with radii of $\sim 1\arcmin$ ($\sim 1$ pc).  Its
brightest optical emission comes from two continuous rings of material
along its northern limb which are distinct in velocity space: A larger
ring of highly redshifted material spanning radial velocities $+5500$
to $0$ \kms, and a smaller ring of mostly blueshifted material
spanning $0$ to $-2500$ \kms.  A handful of additional broken and
complete rings contribute to a velocity asymmetry in the radial
velocities of order $-4500$ to $6000$ \kms.

The observed line profiles of Cas~A's integrated spectrum indicate
that internal absorption obscures or `hides' important kinematic
properties of its ejecta. Given that the integrated profile resulting
from a ring or torus of material suffering no opacity effects is
double-peaked (such that each individual peak is centered at the
approaching and receding velocity edges of the ring), a collection of
co-added double-peaked profiles would be expected from the many ejecta
rings that make up the remnant. Instead, these rings manifest
kinematically in the emission line profiles as line substructure over
top an overall blueshifted distribution. Thus, the combination of
internal absorption with the wide range of orientations, velocities,
and discontinuities of the ejecta rings produces an integrated
emission profile not indicative of its multi-ringed nature.

\begin{figure}[htp!]
\centering
\includegraphics[width=0.75\linewidth]{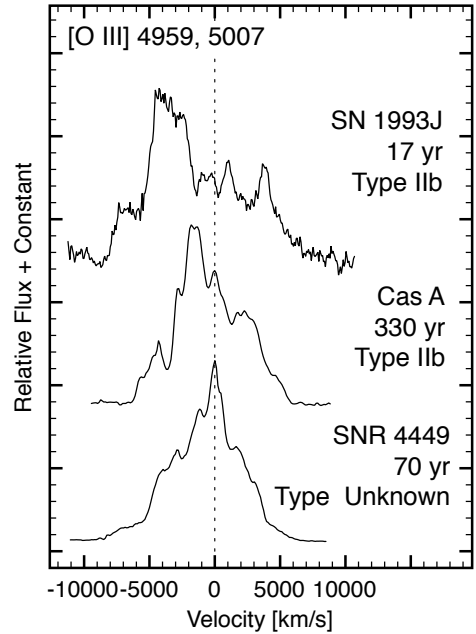}

\caption{[O~III] \dlambda 4959, 5007 emission line profiles of SN~1993J,
  SNR 4449-1, and the integrated Cas~A spectrum. Velocities are with respect to
  5007 \AA. Conspicuous line substructure is seen in all three profiles.}

\label{fig:casa_comparison}
\end{figure}

Emission line substructure in late-time spectra of CCSNe like that
observed in SN 1993J and SNR~4449-1 (Fig.\ \ref{fig:casa_comparison})
is typically identified as `blobs' or `clumps' of ejecta
\citep{Matheson00b,Milisavljevic08a} . However, the multi-peaked line
substructure seen in Cas~A's emissions originate from its multi-ringed
distribution of ejecta. Hence, the similarity between the emission
line profiles of the Cas~A SNR and decades-old CCSNe raises the
possibility that the long-lived emission line substructure of these
extragalactic events may be linked with SN debris arranged in
large-scale rings being excited by a reverse shock like that observed
in Cas~A.

\section{Conclusions}
\label{sec:Conclusions}

Optical {\sl HST} images and ground-based spectra of several CCSNe
obtained years to decades after outburst were investigated in an
effort to understand some of their late-time optical emission
properties. New observations of SN 1957D, 1970G, 1980K, and 1993J were
presented, examined, and compared to archival late-time spectra of six
additional CCSNe retrieved from the literature (SN 1979C, 1986E,
1986J, 1987A, 1996cr, and SNR 4449-1).  Temporal evolution in the
relative strengths and profiles of strong emission lines were
inspected with particular attention to hydrogen and oxygen
emissions.

Properties common to the spectra were found to be consistent with
an association between the long-lived late-time optical emissions and
interaction between SN ejecta and the progenitor star's circumstellar
material. Many objects exhibit conspicuously declining
H$\alpha$/([\ion{O}{1}+[\ion{O}{2}]) and increasing
[\ion{O}{3}]/([\ion{O}{1}]+[\ion{O}{2}]) flux ratios over
time. Narrowing velocity widths of emission lines were also seen in a
handful of cases. These trends are consistent with predictions of
SN--CSM interaction models. The decline of H emission is most likely
due to the H envelope having passed through the reverse shock wave,
and the line narrowing the result of the deceleration of the
interaction shell between the SN ejecta and the CSM at a rate of $\sim 100$
km~s$^{-1}$~yr$^{-1}$.

Asymmetric emission line profiles in oxygen and/or hydrogen emissions
with one or more blueshifted emission peaks were also found to be a
common and long-lasting phenomenon. All spectra in our sample (see
Fig.\ \ref{fig:latetimespec}) exhibit emission profiles having bulk
blueshifts -- but never redshifts -- suggestive of dust formation
within the ejecta or the CDS in the shock region. Many spectra also
show conspicuous line substructure across ionization species.

To further investigate the nature of these emission line asymmetries
and substructures, an integrated optical spectrum of the 330 yr old
Galactic supernova remnant Cas~A was created to simulate what a CCSN
remnant would look like as an unresolved extragalactic
source. Blueshifted profiles with extensive line substructure were
seen in the most prominent lines of [\ion{O}{1}], [\ion{O}{2}],
[\ion{O}{3}], [\ion{S}{2}], and [\ion{Ar}{3}].  The kinematic
properties of the ejecta were mapped with these line profiles. The
multi-peaked line profiles were associated with previously identified
large ring-like structures of high-velocity ejecta, and the lines'
blueshifted velocities were linked to internal absorption resulting
from dust known to reside within the ejecta.

Remarkable similarities are seen between the Cas~A spectrum and
the intermediate-aged CCSN spectra of our sample.  The correspondence
of late-time spectral features between Cas~A and SN 1993J is
particularly strong, which is interesting given that both were Type
IIb events. The shared spectral features are consistent with the view
that emission line asymmetry observed in many evolved CCSN spectra may
be associated with dust in the ejecta. Furthermore, the similarities
between these supernovae suggest that emission line substructures
typically interpreted as ejecta `clumps' or `blobs' in
intermediate-aged CCSNe may actually be linked with large-scale rings
of ejecta as seen in Cas~A.  It is worthwhile to note that Cas~A is
not unique in the spatial distribution of its metal-rich ejecta, as
other young SNRs such as E0102 \citep{Tuohy83,Eriksen01,Vogt10} and
N132D \citep{Morse95,Vogt11} also exhibit ejecta arranged in ring-like
geometries.

\acknowledgements

We thank the referee for comments and suggestions that improved the
manuscript. We also thank John Thorstensen and MDM staff for help with
telescope operations, particularly with configuring the new
acquisition camera which made some of these observations possible, T.\
Matheson and M.\ Modjaz and for making the 2001 spectrum of SN 1993J
available, and F.\ Bauer for providing a spectrum of SN 1996cr. D.M.\
and R.A.F.\ acknowledge support by NSF through grant
AST-0908237. R.A.C.\ acknowledges support from NSF grant AST-0807727.
M.T.\ is supported by the PRIN-INAF 2009 ``Supernovae Variety and
Nucleosynthesis Yields'' and by the grant ASI-INAF I/009/10/0. Based
in part on observations made with the NASA/ESA Hubble Space Telescope,
and obtained from the Hubble Legacy Archive, which is a collaboration
between the Space Telescope Science Institute (STScI/NASA), the Space
Telescope European Coordinating Facility (ST-ECF/ESA) and the Canadian
Astronomy Data Centre (CADC/NRC/CSA). Some data presented here were
obtained at the MMT Observatory, a joint facility of the Smithsonian
Institution and the University of Arizona.  Supernova research at the
Harvard College Observatory is supported by the National Science
Foundation through grants AST-0606772 and AST-0907903.  This research
has made use of SAOImage DS9, developed by Smithsonian Astrophysical
Observatory, and used the facilities of the Canadian Astronomy Data
Centre operated by the National Research Council of Canada with the
support of the Canadian Space Agency.

\end{document}